\documentclass{aa}  

\usepackage{graphicx}
\usepackage{txfonts}
\usepackage{longtable}
\usepackage{subcaption}
\usepackage{lscape}
\usepackage{xcolor}
\def\raa{RAA\ }

\def\cd{\,\mathrm{c/d}}
\def\mmag{\,\mathrm{mmag}}

\begin{document}

   \title{Exploring the unexpected: Disharmonized harmonic pulsation modes in high-amplitude $\delta$ Scuti stars}
\titlerunning{Dis-harmonized harmonics in HADS stars}

   \author{Hui-Fang Xue
   \inst{1,2,3}\fnmsep
   \and Jia-Shu Niu
   \inst{4,5,6}\fnmsep\thanks{Corresponding Author, \email{jsniu@sxu.edu.cn}}
          }

   \institute{Department of Physics, Taiyuan Normal University, Jinzhong 030619, China;\\
             \and
             Institute of Computational and Applied Physics, Taiyuan Normal University, Jinzhong 030619, China;\\
             \and
             Shanxi Key Laboratory for Intelligent Optimization Computing and Blockchain Technology, Jinzhong 030619, China;\\
             \and
             Institute of Theoretical Physics, Shanxi University, Taiyuan 030006, China;\\
        \and
        State Key Laboratory of Quantum Optics and Quantum Optics Devices, Shanxi University, Taiyuan 030006, China;\\
        \and
             Collaborative Innovation Center of Extreme Optics, Shanxi University, Taiyuan, Shanxi 030006, China;\\
             }

   \date{Received XX XX, 2024; accepted XX XX, 2024}

 
\abstract
{Harmonics are a ubiquitous feature across various pulsating stars.\ They are traditionally viewed as mere replicas of the independent primary pulsation modes and have thus been excluded from asteroseismological models. Recent research, however, has uncovered a significant discrepancy: in high-amplitude $\delta$ Scuti (HADS) stars, harmonics exhibit uncorrelated variations in amplitude and frequency relative to their {independent primary pulsation modes}. The nature of these disharmonized harmonics is a question of critical importance. In our study we analysed five triple-mode HADS stars observed by the Transiting Exoplanet Survey Satellite (TESS) and discovered some pervasive patterns of disharmonized harmonics in both the fundamental ($f_0$) and first overtone ($f_1$) pulsation modes. Intriguingly, through an in-depth frequency interaction analysis of V1393 Cen, we identified $2f_1$ as an independent pulsation mode, distinct from $f_1$, and identified it as the progenitor of the frequency variations observed in $3f_1$, $4f_1$, $5f_1$, and $6f_1$. Similar behaviour can be found in DO CMi and GSC 06047-00749, in which $2f_1$ and $3f_1$ are the independent pulsation modes, respectively.
Notably, we found an interesting pattern when decomposing the harmonics that might suggest a generation process of harmonics. These findings serve as a new window on the research of harmonics, which remains a hidden corner of contemporary {asteroseismology}.
}


   \maketitle
%
\section{Introduction}
Harmonics, associated with independent pulsation modes, are a prevalent phenomenon in the constellation of pulsating stars and are present in a wide array of objects, including Cepheids \citep{Rathour2021}, RR Lyrae stars \citep{Kurtz2016}, $\delta$ Scuti stars \citep{Breger2014}, $\gamma$ Dor stars \citep{Kurtz2015}, pulsating white dwarfs \citep{Wu2001}, $\beta$ Cep stars \citep{Degroote2009}, and slowly pulsating B stars \citep{Papics2017}. Traditionally, these harmonics have been ascribed to light curves' non-sinusoidal patterns, a manifestation of the stellar medium's non-linear response to the propagation of pulsation waves. Consequently, they are not classified as intrinsic stellar pulsation modes \citep{Brickhill1992, Wu2001} and are presumed to faithfully replicate the characteristics of their independent parent modes.

Delta Scuti stars, known for their short-period pulsations, have periods ranging from 15 minutes to 8 hours and are classified within spectral types A through F. They are located at the confluence of the main sequence  and the lower boundary of the classical Cepheid instability strip on the Hertzsprung-Russell diagram. The pulsations in these stars are self-excited through the $\kappa$ mechanism, stemming from the partial ionization of helium in the outer layers \citep{Kallinger2008,Handler2009,Guenther2009,Uytterhoeven2011,Holdsworth2014,Steindl2022}.
{The accumulation of high-precision photometric data from space telescopes (such as \textit{Kepler} and TESS) provides the opportunity to study the amplitude and frequency (and phase) variations in $\delta$ Scuti stars in depth, which gives us additional perspectives on the pulsation properties of such stars besides the amplitudes and frequencies \citep[and phases;][]{Bowman2014,Bowman2016,Bowman2016thesis,Guzik2016}}.

The high-amplitude $\delta$ Scuti (HADS) stars  represent a distinct subclass within the $\delta$ Scuti stars, characterized by more pronounced amplitudes ($\Delta V \geq 0_{\cdot}^{m}1$) and slower rotation speeds ($v \sin i \le 30\ \mathrm{km/s}$), although the accumulation of more HADS samples has blurred these traditional distinctions \citep{Balona2012}. The majority of HADS stars oscillate with one or two radial pulsation modes \citep{Niu2013,Niu2017,Xue2018,Bowman2021,Daszynska2022,Xue2022}, yet a subset exhibits more complex pulsation patterns, including three radial modes \citep{Wils2008,Niu2022,Xue2023}, four radial modes \citep{Pietrukowicz2013,Netzel2022a,Netzel2022b}, and even non-radial modes \citep{Poretti2011,Xue2020}.

In the case of the SX Phe star (a Population II subclass of HADS stars) XX Cyg, which features a {spectrogram} dominated by the fundamental pulsation mode ($f_0$) and its 19 harmonics, \citet{Niu2023} discovered that the harmonics exhibit significant and uncorrelated amplitude and frequency variations over time, with these variations becoming more pronounced with increasing harmonic order. Furthermore, for the radial triple-mode HADS star KIC 6382916, the harmonics of the fundamental and first overtone pulsation modes ($f_0$ and $f_1$) display uncorrelated amplitude and frequency variations relative to the {independent primary pulsation modes}, even for the first and second harmonics ($2f_0$, $3f_0$, $2f_1$, and $3f_1$; \citealt{Niu2024}). We refer to these as `disharmonized harmonic pulsation modes', or simply disharmonized harmonics.

These disharmonized harmonics challenge the conventional wisdom regarding the behaviour of harmonics and expose an unexplored aspect of asteroseismology. The question of whether these disharmonized harmonics are a specific feature of certain pulsating stars or represent a widespread phenomenon across certain types of pulsating stars is a critical inquiry in contemporary research.

In this study we scrutinize the amplitude and frequency variations in the harmonics of five HADS stars: DO CMi, GSC 06047-00749 (ASAS J094303-1707.3), V0803 Aur, V1384 Tau, and V1393 Cen. These stars have been identified as radial triple-mode HADS stars based on continuous time-series photometric data from the Transiting Exoplanet Survey Satellite (TESS) space telescope \citep{Xue2023}, {whose amplitudes and frequencies for the fundamental, first overtone, and second overtone pulsation modes ($A_0$, $A_1$, $A_2$ and $f_0$, $f_1$, $f_2$) are presented in Table \ref{tab:basic_info}; the spectrograms of the five HADS stars are presented in Fig. \ref{fig:spectra}.}

\begin{table*}[hbtp!]
\centering
  \caption{Amplitudes and frequencies of the fundamental, first overtone, and second overtone pulsation modes of the five triple-mode HADS stars.}
  \label{tab:basic_info}
  \resizebox{1.0\textwidth}{!}{
  \begin{tabular}{l|c|c|c|c|c}
    \hline
    \hline
  ID & DO CMi & GSC 06047-00749 & V0803 Aur & V1384 Tau & V1393 Cen \\
    \hline
  $A_0$ ($\mmag$) & $49\pm 1$ & $53\pm 1$ & $49.5\pm 0.6$ & $52.8\pm 0.8$ & $41.3\pm 0.5$ \\ 
  $f_0$ ($\cd$)   & $5.1416\pm 0.0005$ & $10.0829\pm 0.0004$ & $14.0735\pm 0.0003$ & $7.1534\pm 0.0003$ & $8.4902\pm 0.0003$ \\ 
  $A_1$ ($\mmag$) & $49.0\pm 0.7$ & $61\pm 1$ & $20.8\pm 0.2$ & $49.9\pm 0.5$ & $71\pm 1$ \\ 
  $f_1$ ($\cd$)   & $6.7284\pm 0.0003$ & $13.0690\pm 0.0005$ & $18.1716\pm 0.0003$ & $9.3117\pm 0.0002$ & $11.0093\pm 0.0003$ \\ 
  $A_2$ ($\mmag$) & $11.6\pm 0.2$ & $1.42\pm 0.07$ & $4.5\pm 0.1$ & $4.41\pm 0.07$ & $5.65\pm $  0.09 \\ 
  $f_2$ ($\cd$)   & $8.6078\pm 0.0004$ & $16.303\pm 0.001$ & $22.5272\pm 0.0005$ & $11.6430\pm 0.0003$ & $13.7271\pm 0.0003$ \\
  \hline
 & {07$^{d}$ (24 d), 33$^{*,c}$ (26 d)} & {08$^{a}$ (24 d), 35$^{a}$ (24 d),} & {43$^{*,a}$ (24 d), 44$^{a}$ (24 d),} & {05$^{*,a}$ (25 d), 70$^{b}$ (25 d),} & {11$^{a}$ (25 d), 38$^{*,a}$ (27d), } \\
  TESS Sector  & & {62$^{*,b}$ (24 d)} &{45$^{a}$ (25 d), 60$^{b}$ (26 d),} & {71$^{b}$ (26 d)} & {65$^{b}$ (27 d)}\\
  & &  &{71$^{b}$ (26 d), 72$^{b}$ (25 d)} & & \\
\hline
\end{tabular}
}
\\
\footnotesize{Note: $^{*}$ denotes the data used in this work; $^{a}$, $^{b}$, $^{c}$, and $^{d}$ denote the data cadence of 120 s, 200 s, 600 s, and 1800 s, respectively; the number in parentheses denotes the length of the dataset in days.}
\end{table*}

\section{Results }
Ultimately, we extracted the variation data (amplitudes and frequencies) for the fundamental and first overtone pulsation modes and their harmonics (denoted as $mf_0$ and $nf_1$, where $m, n \in \mathbb{N}$ and $m, n > 1$) for the five HADS stars (see Appendix \ref{app:method} for more details on the data reduction, {and Appendix \ref{app:SM} for a demonstration of the validity of the methods}). The precise number of harmonics was contingent upon the signal-to-noise ratio within the moving time windows, with no significant signals detected for the harmonics of the second overtone pulsation mode. 

\begin{figure*}[htp]
  \centering
  \includegraphics[width=0.48\textwidth]{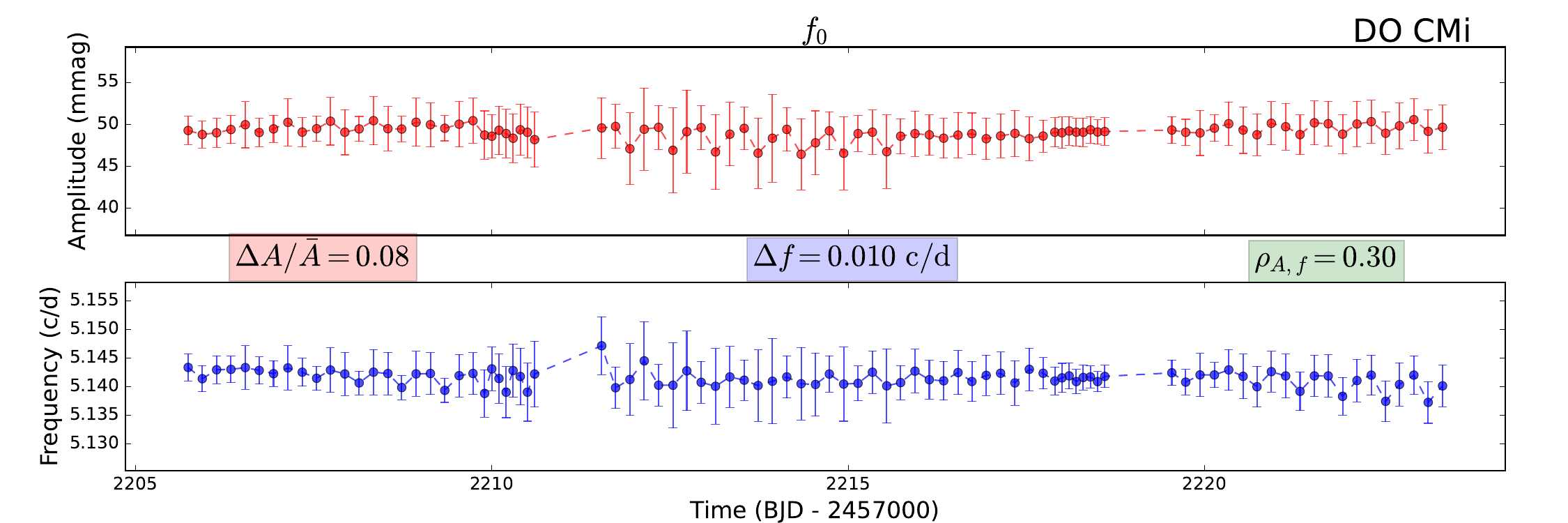}
  \includegraphics[width=0.48\textwidth]{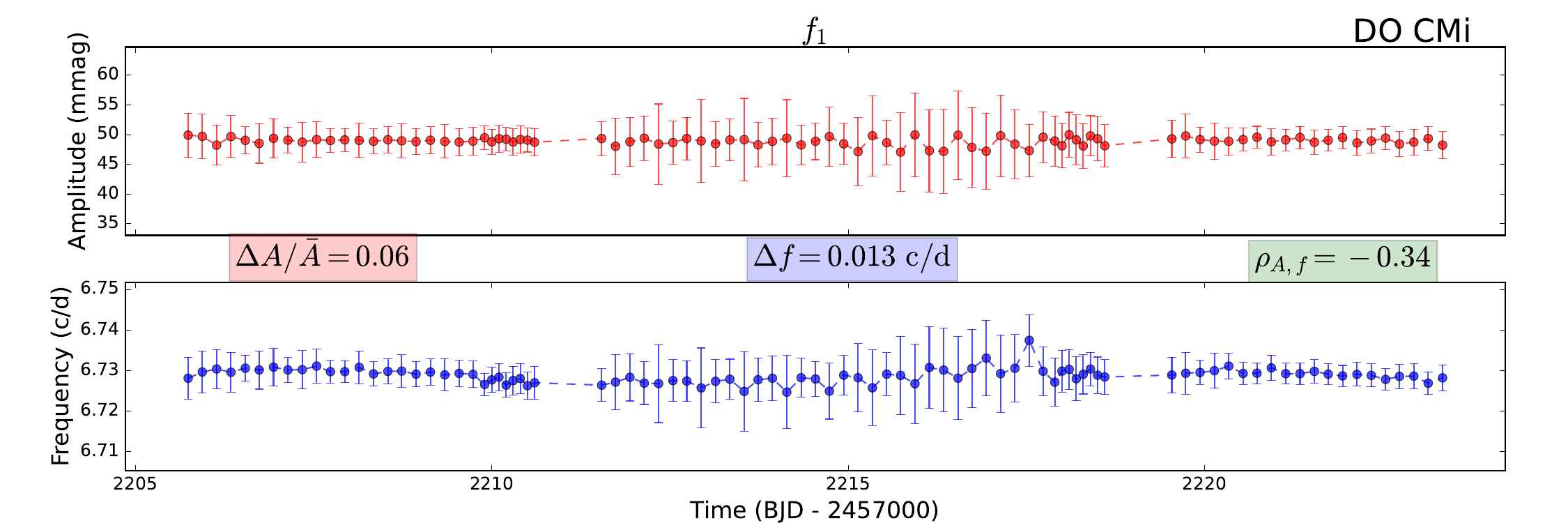}
  \includegraphics[width=0.48\textwidth]{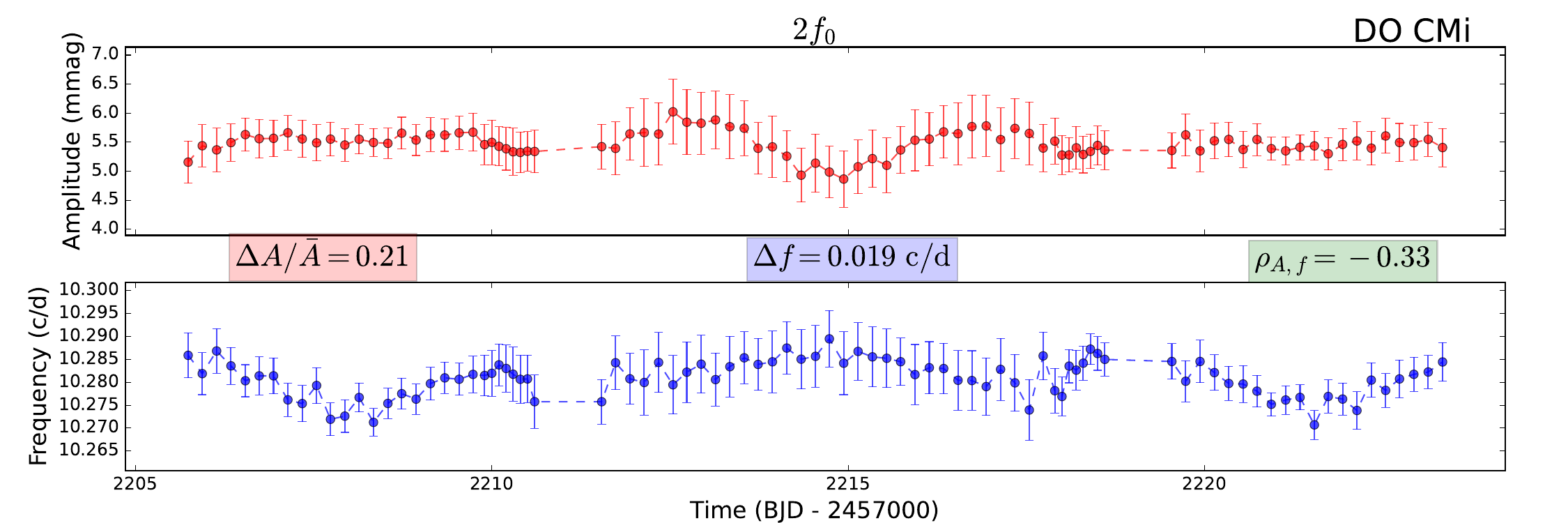}
  \includegraphics[width=0.48\textwidth]{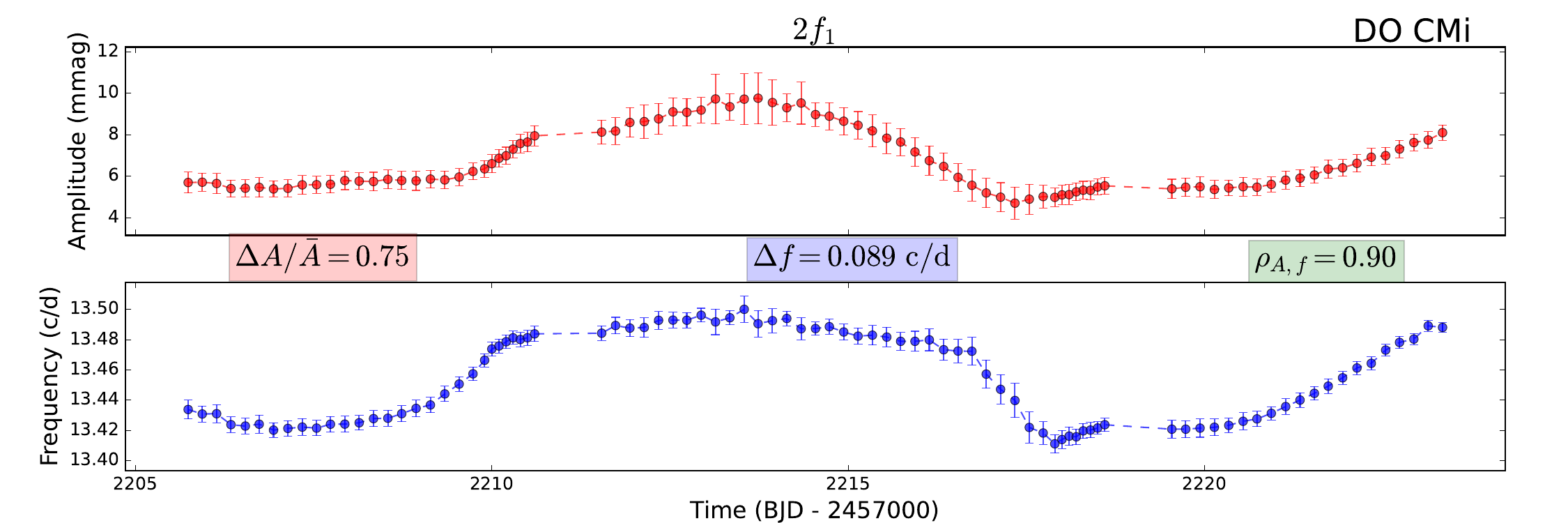}
  \includegraphics[width=0.48\textwidth]{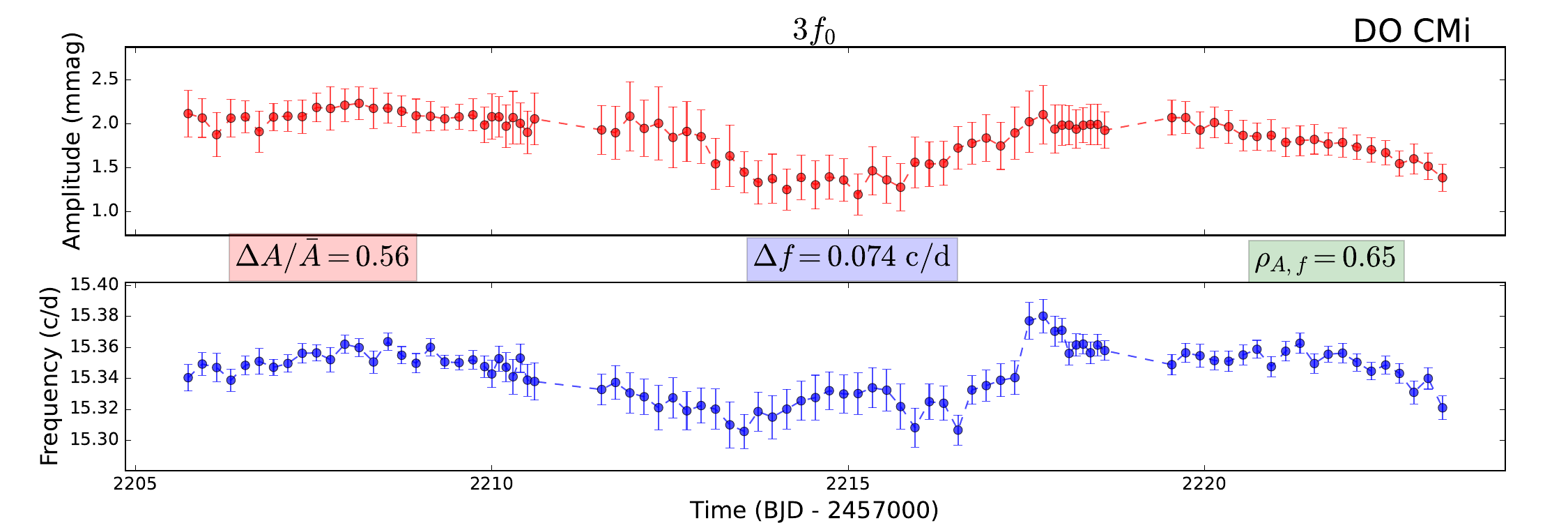}
  \includegraphics[width=0.48\textwidth]{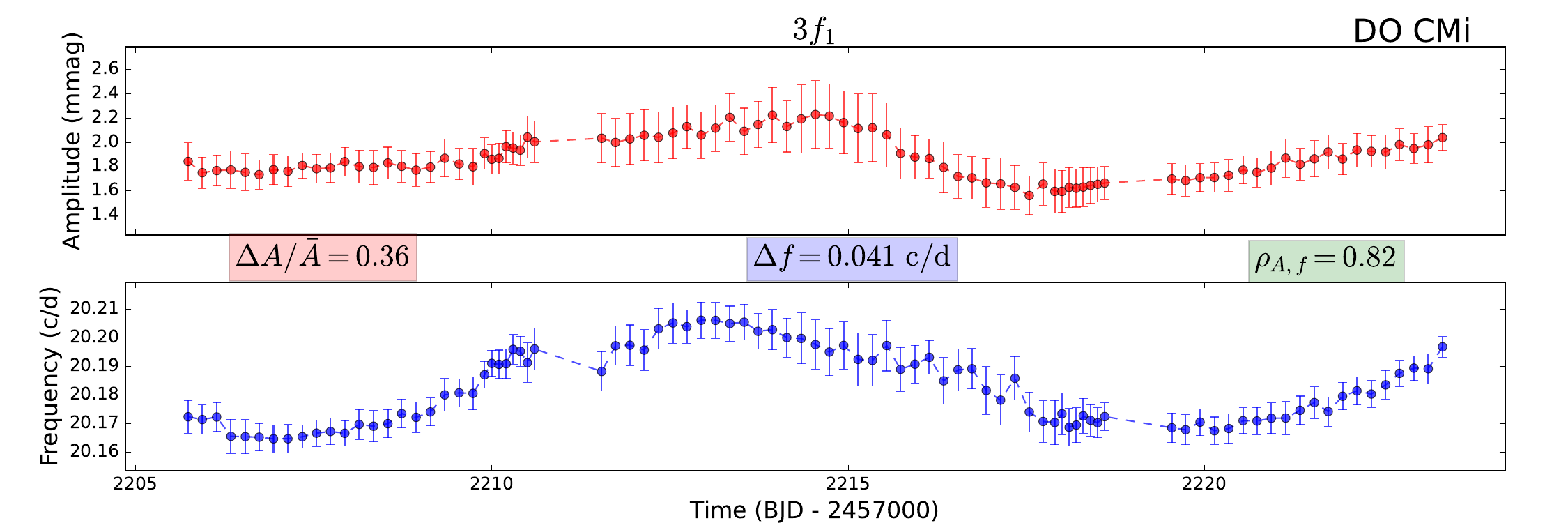}
  \caption{Amplitude and frequency variations in the fundamental and first overtones and their harmonic pulsation modes ({$f_{0}$ through $3f_{0}$} and {$f_{1}$ through $3f_{1}$}) in DO CMi.}
  \label{fig:DOCMi}
\end{figure*}

\begin{figure*}[htp]
  \centering
  \includegraphics[width=0.48\textwidth]{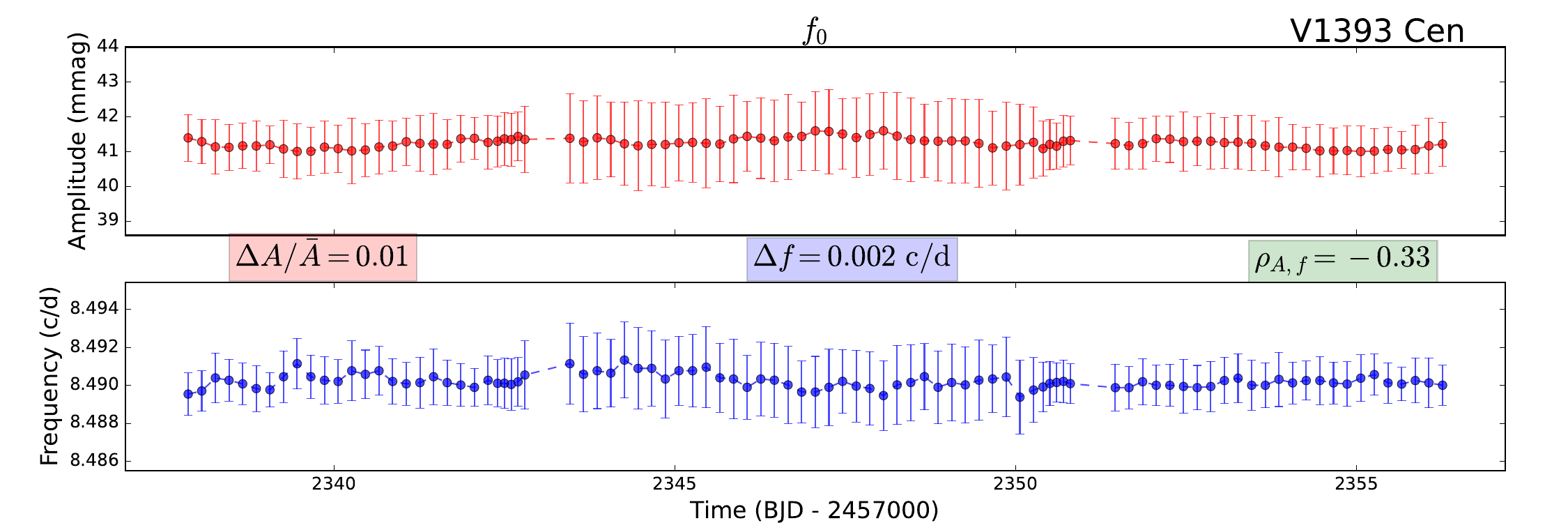}
  \includegraphics[width=0.48\textwidth]{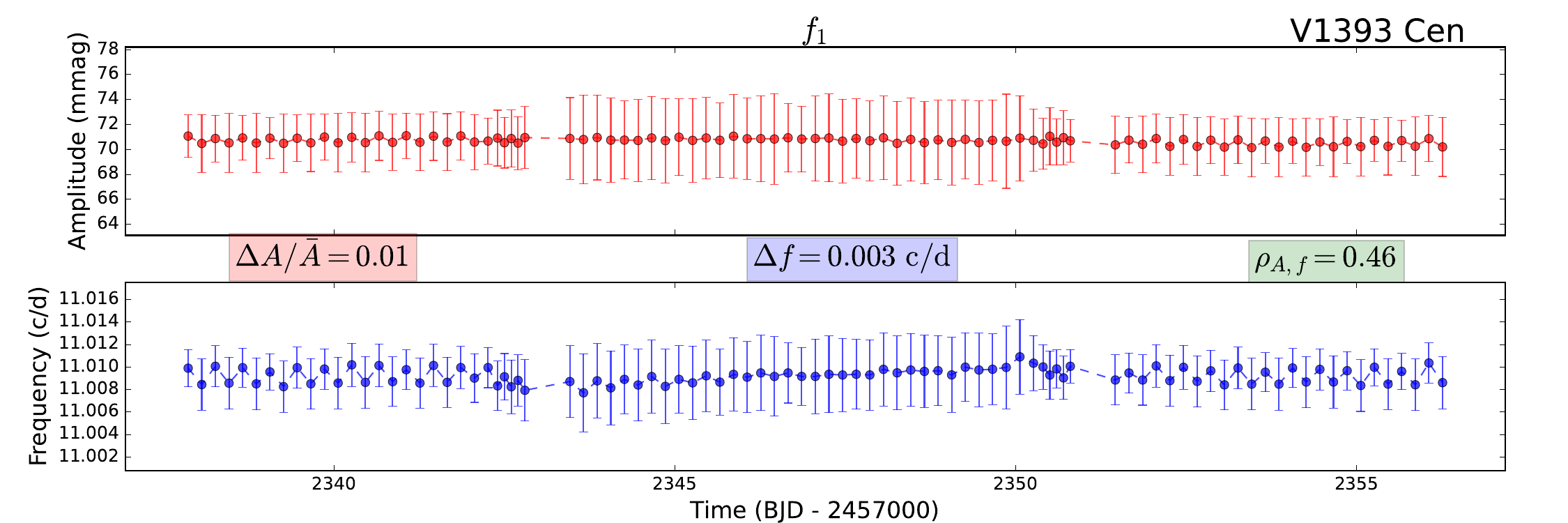}
  \includegraphics[width=0.48\textwidth]{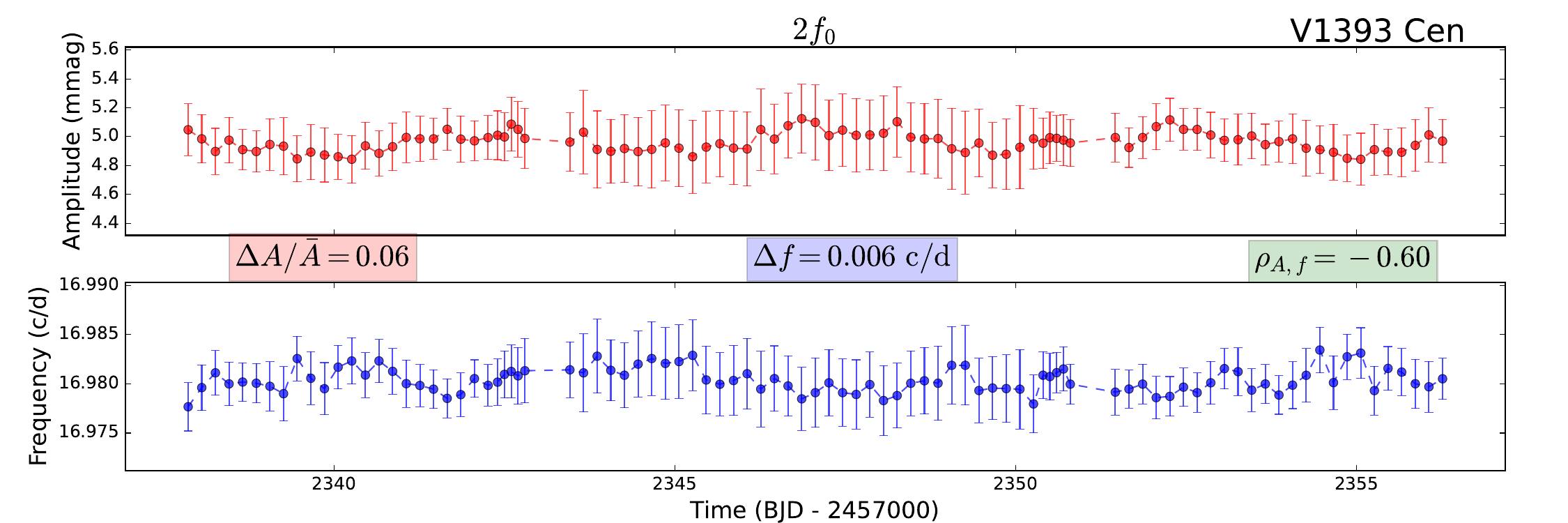}
  \includegraphics[width=0.48\textwidth]{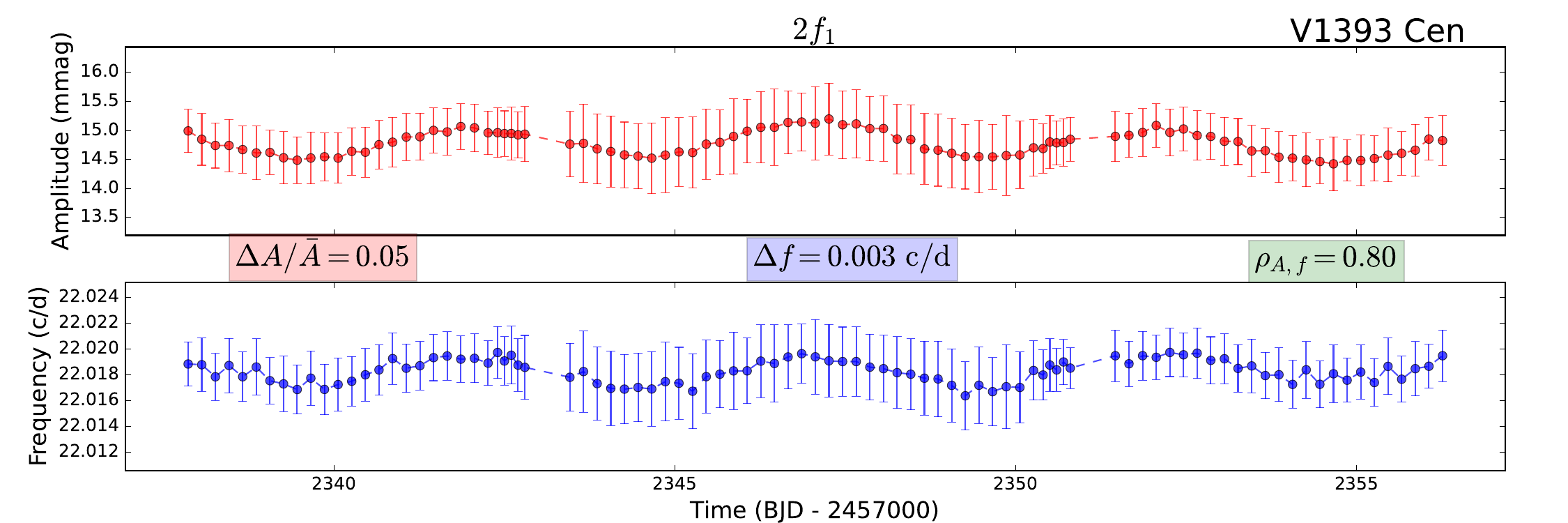}
  \includegraphics[width=0.48\textwidth]{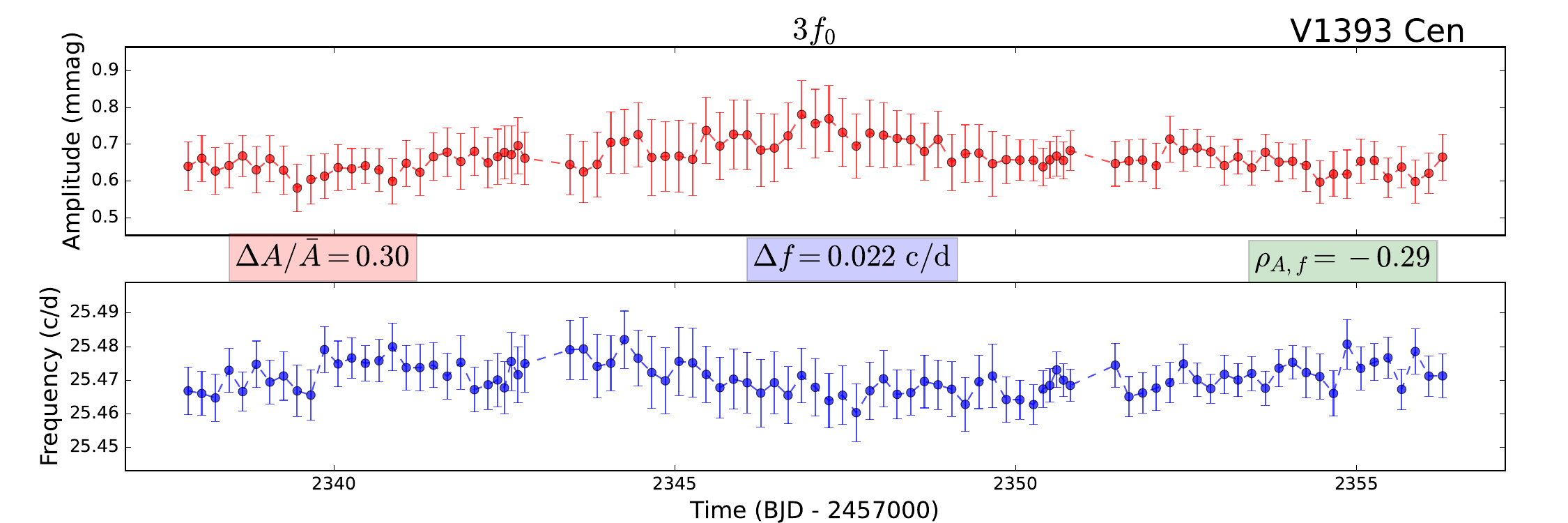}
  \includegraphics[width=0.48\textwidth]{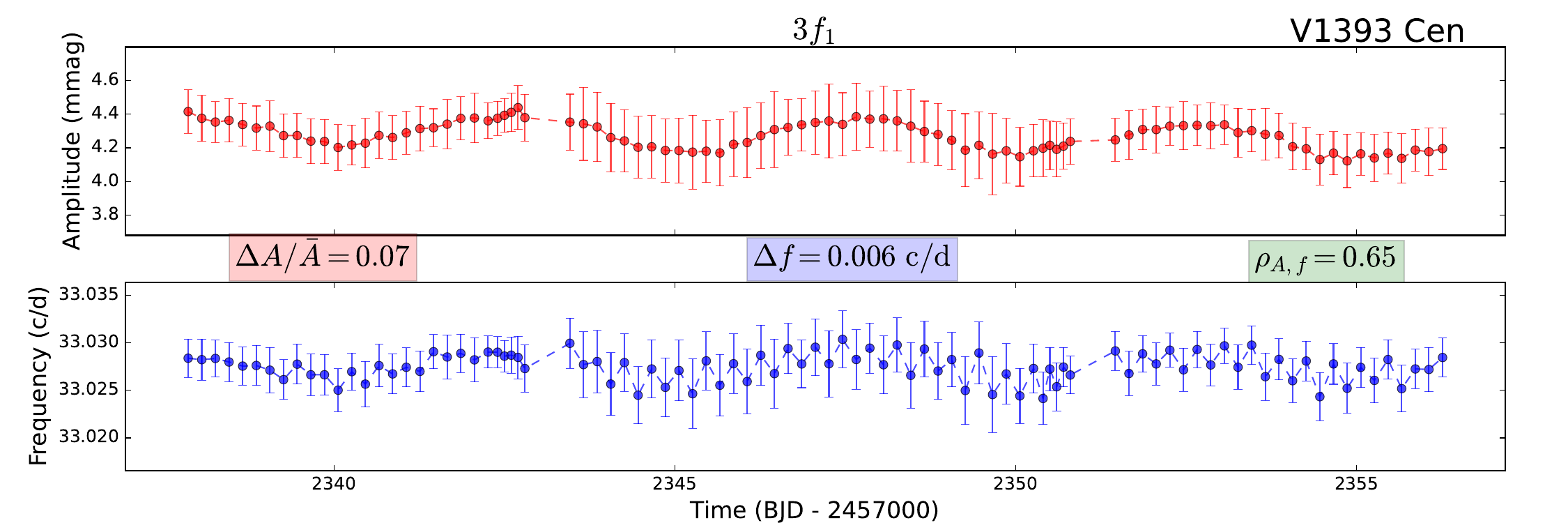}
  \phantom{\includegraphics[width=0.48\textwidth]{fig/V1393Cen_3f0.pdf}}
  \includegraphics[width=0.48\textwidth]{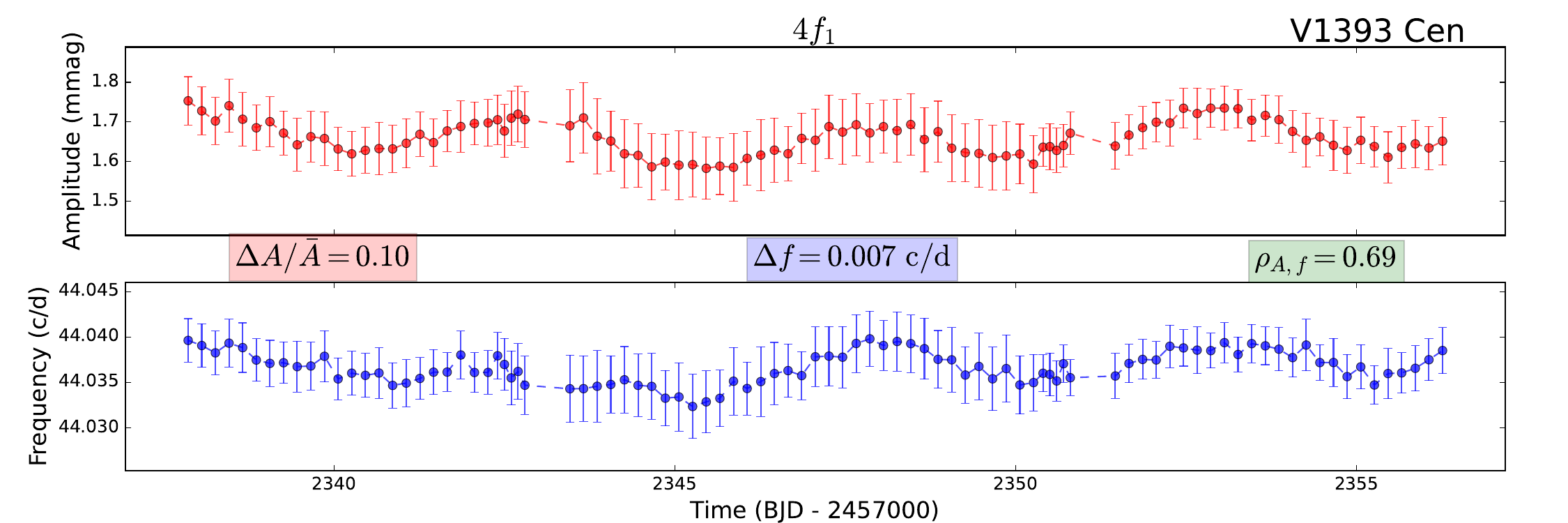}
  \phantom{\includegraphics[width=0.48\textwidth]{fig/V1393Cen_3f0.pdf}}
  \includegraphics[width=0.48\textwidth]{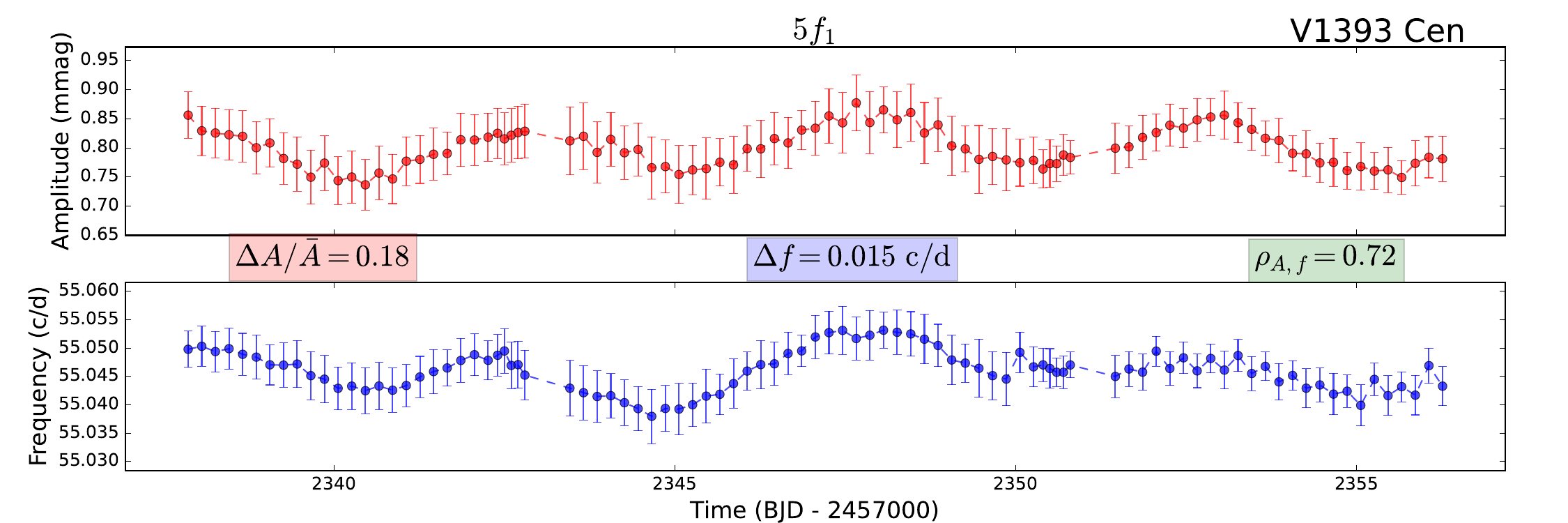}
  \phantom{\includegraphics[width=0.48\textwidth]{fig/V1393Cen_3f0.pdf}}
  \includegraphics[width=0.48\textwidth]{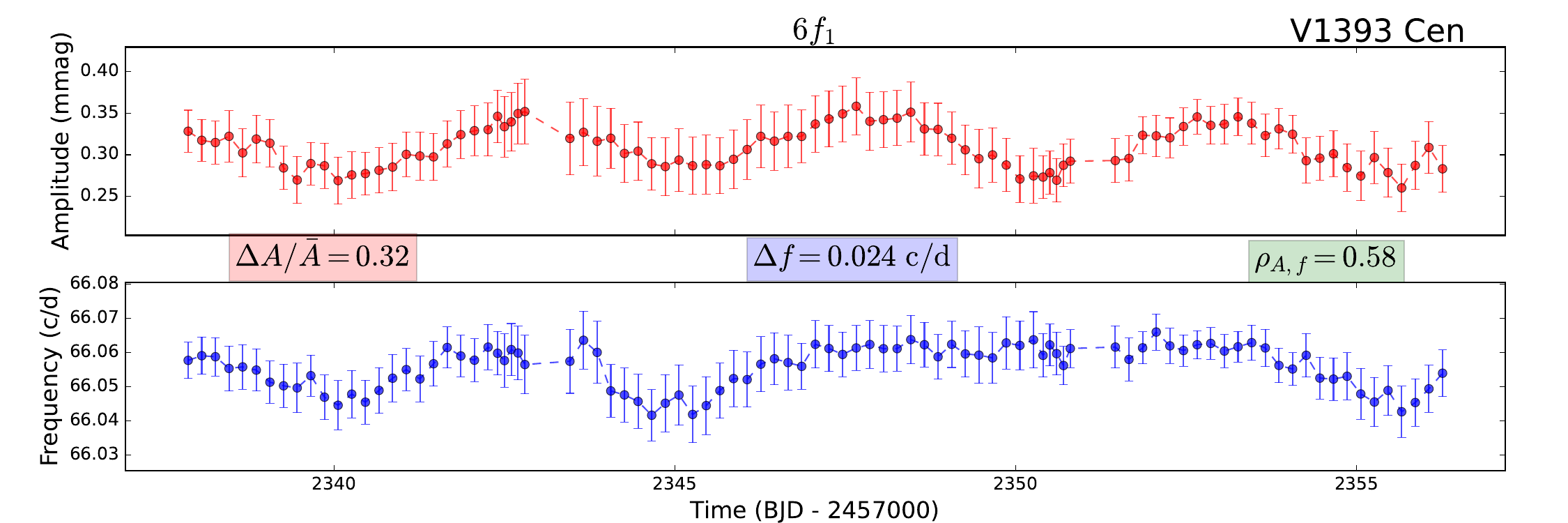}%
  \caption{Amplitude and frequency variations in the fundamental and first overtones and their harmonic pulsation modes ({$f_{0}$ through $3f_{0}$} and {$f_{1}$ through $6f_{1}$}) in V1393 Cen.}
  \label{fig:V1393Cen}
\end{figure*}

The variations in amplitude and frequency for the fundamental and first overtones and their associated harmonic pulsation modes in DO CMi and V1393 Cen are depicted in Figs. \ref{fig:DOCMi} (DO CMi: {$f_{0}$ through $3f_{0}$} and {$f_{1}$ through $3f_{1}$}) and \ref{fig:V1393Cen} (V1393 Cen: {$f_{0}$ through $3f_{0}$} and {$f_{1}$ through $6f_{1}$}), respectively. {They both show evident correlated amplitude and frequency  modulations in some of their harmonics}. The corresponding data for the remaining three HADS stars are presented in Figs. \ref{fig:GSC06047-00749} (GSC 06047-00749: {$f_{0}$ through $3f_{0}$} and {$f_{1}$ through $4f_{1}$}), \ref{fig:V0803Aur} (V0803 Aur: {$f_{0}$ through $4f_{0}$} and {$f_{1}$ through $2f_{1}$}), and \ref{fig:V1384Tau} (V1384 Tau: {$f_{0}$ through $3f_{0}$} and {$f_{1}$ through $3f_{1}$}).

To quantitatively illustrate the variations in amplitude and frequency, as well as their interrelations, we calculated the relative amplitude variations ($\Delta A/\bar{A} \equiv (A_\mathrm{max} - A_\mathrm{min})/\bar{A}$)\footnote{Here, $A_\mathrm{max}$ and $A_\mathrm{min}$ represent the maximum and minimum amplitude values within the moving windows, while $\bar{A}$ denotes the mean amplitude across these windows.}, the absolute frequency variations ($\Delta f \equiv f_\mathrm{max} - f_\mathrm{min}$)\footnote{$f_\mathrm{max}$ and $f_\mathrm{min}$ correspond to the maximum and minimum frequency values observed within the moving windows.}, and the Pearson correlation coefficient ($\rho_{A,f}$) between amplitude and frequency. These quantitative metrics are displayed in Figs. \ref{fig:DOCMi}, \ref{fig:V1393Cen}, \ref{fig:GSC06047-00749}, \ref{fig:V0803Aur}, and \ref{fig:V1384Tau}, providing a detailed account of the observed phenomena.

\section{Conclusions and discussions}

Typically, the amplitudes and frequencies of the fundamental ($f_0$) and first overtone ($f_1$) pulsation modes exhibit a high degree of stability over a timescale of more than ten days. However, two distinct behaviours appear when the order of the harmonics increases: (a) a fluctuation (which deviates from the stability of $f_0$ and $f_1$) grows progressively with the increasing order; and (b) a marked modulation is abruptly injected into a harmonic of a certain order, and  all the higher-order harmonics then present  amplitude and frequency variations that are highly correlated to it. 
The former seems to be a general trend in all five HADS stars and is similar to the case of XX Cyg \citep{Niu2023}. The latter is only evident in some cases (such as $2f_1$ and $3f_1$ in DO CMi, $3f_1$ and $4f_1$ in GSC 06047-00749, and $2f_1$ through $6f_1$ in V1393 Cen), which is similar to the case of KIC 6382916 \citep{Niu2024}.
Both behaviours indicate that the disharmonized harmonics are commonly encountered in triple-mode HADS stars and suggest the presence of as yet unidentified mechanisms

On a shorter timescale of approximately 0.2 to 1.0 days, minor frequency modulations become apparent, affecting certain pulsation modes. This effect is notably evident in DO CMi ($f_0$ and $f_1$), V1393 Cen ($f_1$ and $3f_1$), GSC 06047-00749 ($f_1$), and V1384 Tau ($f_0$). The observation of a similar 0.6570-day modulation in the time-frequency diagram of KIC 6382916 \citep{Niu2024} indicates that such modulations may be a common characteristic of triple-mode HADS stars, although the underlying physical processes remain elusive.

The disharmonized harmonics exhibit distinct behaviours for $f_0$ and $f_1$. For $f_0$, the amplitude and frequency variations in its harmonics do not exhibit a strong correlation; this also what is seen with $2f_0$ and $3f_0$ in DO CMi. In contrast, for $f_1$, there is a significant correlation in the amplitude and frequency variations, which is particularly evident in DO CMi ($2f_1$ and $3f_1$) and V1393 Cen ($2f_1$, $3f_1$, $4f_1$, $5f_1$, and $6f_1$).

An intriguing observation is the significant (anti-)correlated relationships in the variations in amplitude and frequency for certain pulsation modes, as precisely quantified by the Pearson correlation coefficient ($\rho_{A,f}$) and presented in the figures. This is notably observed in DO CMi ($3f_0$, $2f_1$, and $3f_1$), V1393 Cen ($2f_0$, $2f_1$, $3f_1$, $4f_1$, $5f_1$, and $6f_1$), and GSC 06047-00749 ($3f_1$ and $4f_1$). These (anti-)correlations are significant as they suggest that the observed amplitude and frequency variations may be attributed to modulations resulting from interactions between pulsation modes \citep{Niu2024}.

\begin{figure*}[htp]
  \centering
  \includegraphics[width=0.7\textwidth]{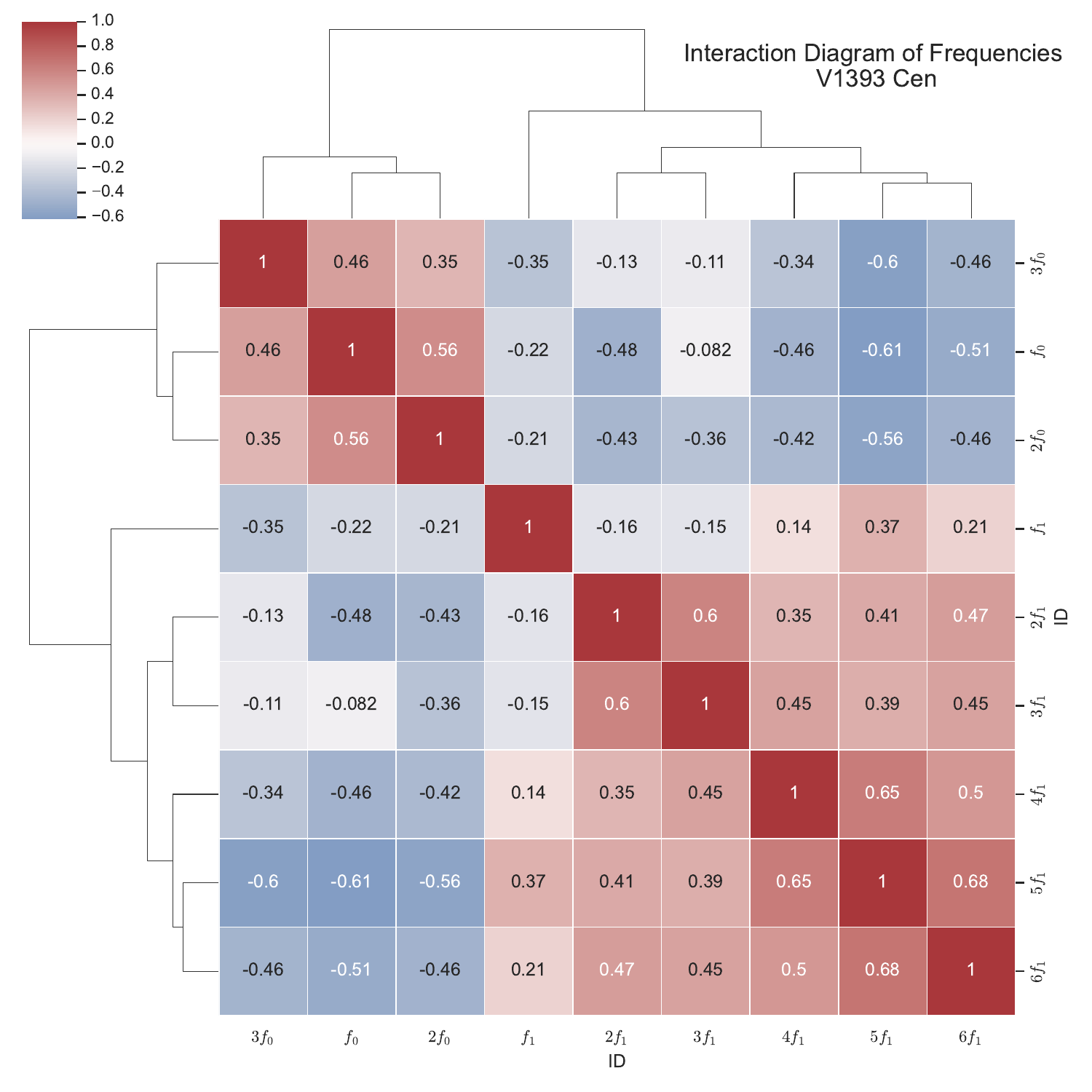}
  \caption{Interaction diagram of the frequencies of $f_0$ and $f_1$ and their harmonic pulsation modes in V1393 Cen.}
  \label{fig:V1393Cen_ID}
\end{figure*}

The harmonics of $f_1$ in V1393 Cen, which display a high degree of correlation, provide a valuable context for exploring the detailed relationships among all $f_0$ and $f_1$ harmonics. We analysed these harmonics by constructing an interaction diagram of the frequencies (Fig. \ref{fig:V1393Cen_ID})\footnote{For the subsequent quantitative analysis, the correlation coefficients between frequency variations are given in the coloured squares for reference. {The interaction diagrams of frequencies for the other four stars are presented in Appendix \ref{app:IDs}.}}. The frequency variations in the pulsation modes are grouped into two distinct categories: those related to $f_0$ and those related to $f_1$. As depicted in Fig. \ref{fig:V1393Cen_ID}, all $f_1$-related harmonics, with the exception of $f_1$ itself, exhibit significant correlations. This is analogous to the findings for KIC 6382916 \citep{Niu2024}, where $2f_1$ is considered an independent parent mode for $3f_1$ due to their significant correlation. The correct representation of $3f_1$ should thus be $(2f_1) + (f_1)$ rather than $3 \cdot (f_1)$.\footnote{The parentheses here indicate that the enclosed pulsation modes should be regarded as a collective entity rather than mere multiples of $f_1$.}

Applying the principle that a harmonic's parent mode should exhibit the highest correlation in frequency variations, we can deduce the correct representation of all harmonics as follows:
\begin{align}
  (6f_1) &= (5f_1) + (f_1), \notag \\
  (5f_1) &= (4f_1) + (f_1), \notag \\
  (4f_1) &= (3f_1) + (f_1), \notag \\
  (3f_1) &= (2f_1) + (f_1). 
  \label{eq:6-3f1}
\end{align}
Each of these high-order harmonics ($nf_1$, $n \ge 3$) have two parent modes: $f_1$ and $(n-1)f_1$ (the latter passes on its variation patterns to $nf_1$). 
Even more interesting from a global perspective is that these higher-order harmonics seem to be generated step by step, that is to say, $nf_1$ is directly generated by $(n-1)f_1$ rather than the lower-order harmonics. 
This interesting generation mechanism of harmonics explains the ubiquitous phenomenon in the spectrogram of many pulsating stars: harmonics appear one by one in order, and their amplitudes decrease as the order increases.

However, the relationship
\begin{equation}
  (2f_1) \neq (f_1) + (f_1)
  \label{eq:2f1}
\end{equation}
holds true due to the negative correlation coefficient ($-0.16$) between $f_1$ and $2f_1$. Consequently, considering the frequency variations, $2f_1$ emerges as an independent pulsation mode, the progenitor of the significant frequency variations observed in higher-order harmonics. This pattern is also evident in KIC 6382916 \citep{Niu2024} and DO CMi (as shown in Fig. \ref{fig:DOCMi}). {Moreover, a similar case also occurs in GSC 06047-00749, but with a different relationship: $(3f_1) \neq (f_1) + (2f_1)$ and $(3f_1) \neq (f_1) + (f_1) + (f_1)$.}

The origin of the frequency variations in $2f_1$ needs further investigation and may be linked to the nature of $f_2$ as a resonating integration mode, as well as the resonance relationship between $f_0$, $f_1$, and $f_2$ as expressed by the equation
\begin{align}
  2f_1 - f_0 = f_2 - \Delta \omega,
\label{eq:resonance}
\end{align}
where $\Delta \omega$ represents the star's rotational frequency. Elucidating the details of this relationship is an area earmarked for future research.

In summary, the present study has provided compelling evidence for the prevalence of disharmonized harmonics in HADS stars, challenging the traditional paradigm of harmonic behaviour in pulsating stars. The detailed analysis of five triple-mode HADS stars has unveiled a complex pattern of amplitude and frequency variations that cannot be solely attributed to the {independent primary pulsation modes}. The discovery of $2f_1$ as an independent pulsation mode, with its own intrinsic frequency variations, suggests the existence of a multifaceted mechanism governing the pulsations. An intuitive but non-trivial discovery suggests a generation process of harmonics. These findings call for a reassessment of the current models and invite further investigation into the underlying physics, which would potentially lead to a more comprehensive understanding of the pulsation phenomena in $\delta$ Scuti stars and other variable stars.

\begin{acknowledgements}
We would like to thank Jue-Ran Niu for providing us with an efficient working environment. H.F.X. acknowledges support from the National Natural Science Foundation of China (NSFC) (No. 12303036) and the Applied Basic Research Programs of Natural Science Foundation of Shanxi Province (No. 202103021223320). J.S.N. acknowledges support from the National Natural Science Foundation of China (NSFC) (No. 12005124). 
The authors acknowledge the TESS Science team and everyone who has contributed to making the TESS mission possible. 
\end{acknowledgements}

%
%

\begin{thebibliography}{38}
\expandafter\ifx\csname natexlab\endcsname\relax\def\natexlab#1{#1}\fi

\bibitem[{Aerts(2021)}]{Aerts2021}
Aerts, C. 2021, Rev. Mod. Phys., 93, 015001

\bibitem[{{Balona} {et~al.}(2012){Balona}, {Lenz}, {Antoci}, {Bernabei},
  {Catanzaro}, {Daszy{\'n}ska-Daszkiewicz}, {di Criscienzo}, {Grigahc{\`e}ne},
  {Handler}, {Kurtz}, {Marconi}, {Molenda-{\.Z}akowicz}, {Moya}, {Nemec},
  {Pigulski}, {Pricopi}, {Ripepi}, {Smalley}, {Su{\'a}rez}, {Suran}, {Hall},
  {Kinemuchi}, \& {Klaus}}]{Balona2012}
{Balona}, L.~A., {Lenz}, P., {Antoci}, V., {et~al.} 2012, \mnras, 419, 3028

\bibitem[{{Bowman}(2016)}]{Bowman2016thesis}
{Bowman}, D.~M. 2016, PhD thesis, University of Central Lancashire, UK

\bibitem[{{Bowman} {et~al.}(2021){Bowman}, {Hermans},
  {Daszy{\'n}ska-Daszkiewicz}, {Holdsworth}, {Tkachenko}, {Murphy}, {Smalley},
  \& {Kurtz}}]{Bowman2021}
{Bowman}, D.~M., {Hermans}, J., {Daszy{\'n}ska-Daszkiewicz}, J., {et~al.} 2021,
  \mnras, 504, 4039

\bibitem[{{Bowman} \& {Kurtz}(2014)}]{Bowman2014}
{Bowman}, D.~M. \& {Kurtz}, D.~W. 2014, \mnras, 444, 1909

\bibitem[{{Bowman} {et~al.}(2016){Bowman}, {Kurtz}, {Breger}, {Murphy}, \&
  {Holdsworth}}]{Bowman2016}
{Bowman}, D.~M., {Kurtz}, D.~W., {Breger}, M., {Murphy}, S.~J., \&
  {Holdsworth}, D.~L. 2016, \mnras, 460, 1970

\bibitem[{{Breger} \& {Montgomery}(2014)}]{Breger2014}
{Breger}, M. \& {Montgomery}, M.~H. 2014, \apj, 783, 89

\bibitem[{{Brickhill}(1992)}]{Brickhill1992}
{Brickhill}, A.~J. 1992, \mnras, 259, 519

\bibitem[{{Daszy{\'n}ska-Daszkiewicz}
  {et~al.}(2022){Daszy{\'n}ska-Daszkiewicz}, {Walczak}, {Pamyatnykh}, \&
  {Szewczuk}}]{Daszynska2022}
{Daszy{\'n}ska-Daszkiewicz}, J., {Walczak}, P., {Pamyatnykh}, A.~A., \&
  {Szewczuk}, W. 2022, \mnras, 512, 3551

\bibitem[{{Degroote} {et~al.}(2009){Degroote}, {Briquet}, {Catala},
  {Uytterhoeven}, {Lefever}, {Morel}, {Aerts}, {Carrier}, {Auvergne}, {Baglin},
  \& {Michel}}]{Degroote2009}
{Degroote}, P., {Briquet}, M., {Catala}, C., {et~al.} 2009, \aap, 506, 111

\bibitem[{{Guenther} {et~al.}(2009){Guenther}, {Kallinger}, {Zwintz}, {Weiss},
  {Kuschnig}, {Casey}, {Matthews}, {Moffat}, {Rucinski}, {Sasselov}, \&
  {Walker}}]{Guenther2009}
{Guenther}, D.~B., {Kallinger}, T., {Zwintz}, K., {et~al.} 2009, \apj, 704,
  1710

\bibitem[{{Guzik} {et~al.}(2016){Guzik}, {Kosak}, {Bradley}, \&
  {Jackiewicz}}]{Guzik2016}
{Guzik}, J.~A., {Kosak}, K., {Bradley}, P.~A., \& {Jackiewicz}, J. 2016, IAU
  Focus Meeting, 29B, 560

\bibitem[{{Handler}(2009)}]{Handler2009}
{Handler}, G. 2009, in American Institute of Physics Conference Series, Vol.
  1170, Stellar Pulsation: Challenges for Theory and Observation, ed. J.~A.
  {Guzik} \& P.~A. {Bradley}, 403--409

\bibitem[{{Holdsworth} {et~al.}(2014){Holdsworth}, {Smalley}, {Gillon},
  {Clubb}, {Southworth}, {Maxted}, {Anderson}, {Barros}, {Collier Cameron},
  {Delrez}, {Faedi}, {Haswell}, {Hellier}, {Horne}, {Jehin}, {Norton},
  {Pollacco}, {Skillen}, {Smith}, {West}, \& {Wheatley}}]{Holdsworth2014}
{Holdsworth}, D.~L., {Smalley}, B., {Gillon}, M., {et~al.} 2014, \mnras, 439,
  2078

\bibitem[{{Jenkins} {et~al.}(2016){Jenkins}, {Twicken}, {McCauliff},
  {Campbell}, {Sanderfer}, {Lung}, {Mansouri-Samani}, {Girouard}, {Tenenbaum},
  {Klaus}, {Smith}, {Caldwell}, {Chacon}, {Henze}, {Heiges}, {Latham},
  {Morgan}, {Swade}, {Rinehart}, \& {Vanderspek}}]{Jenkins2016}
{Jenkins}, J.~M., {Twicken}, J.~D., {McCauliff}, S., {et~al.} 2016, in Society
  of Photo-Optical Instrumentation Engineers (SPIE) Conference Series, Vol.
  9913, Software and Cyberinfrastructure for Astronomy IV, ed. G.~{Chiozzi} \&
  J.~C. {Guzman}, 99133E

\bibitem[{{Kallinger} {et~al.}(2008){Kallinger}, {Zwintz}, \&
  {Weiss}}]{Kallinger2008}
{Kallinger}, T., {Zwintz}, K., \& {Weiss}, W. 2008, \aap, 488, 279

\bibitem[{{Kurtz} {et~al.}(2016){Kurtz}, {Bowman}, {Ebo}, {Moskalik},
  {Handberg}, \& {Lund}}]{Kurtz2016}
{Kurtz}, D.~W., {Bowman}, D.~M., {Ebo}, S.~J., {et~al.} 2016, \mnras, 455, 1237

\bibitem[{{Kurtz} {et~al.}(2015){Kurtz}, {Shibahashi}, {Murphy}, {Bedding}, \&
  {Bowman}}]{Kurtz2015}
{Kurtz}, D.~W., {Shibahashi}, H., {Murphy}, S.~J., {Bedding}, T.~R., \&
  {Bowman}, D.~M. 2015, \mnras, 450, 3015

\bibitem[{{Montgomery} \& {O'Donoghue}(1999)}]{Montgomery1999}
{Montgomery}, M.~H. \& {O'Donoghue}, D. 1999, Delta Scuti Star Newsletter, 13,
  28

\bibitem[{{Netzel} {et~al.}(2022){Netzel}, {Pietrukowicz}, {Soszy{\'n}ski}, \&
  {Wrona}}]{Netzel2022a}
{Netzel}, H., {Pietrukowicz}, P., {Soszy{\'n}ski}, I., \& {Wrona}, M. 2022,
  \mnras, 510, 1748

\bibitem[{{Netzel} \& {Smolec}(2022)}]{Netzel2022b}
{Netzel}, H. \& {Smolec}, R. 2022, \mnras, 515, 4574

\bibitem[{{Niu} {et~al.}(2017){Niu}, {Fu}, {Li}, {Yang}, {Zong}, {Xue},
  {Zhang}, {Liu}, {Du}, \& {Zuo}}]{Niu2017}
{Niu}, J.-S., {Fu}, J.-N., {Li}, Y., {et~al.} 2017, \mnras, 467, 3122

\bibitem[{{Niu} {et~al.}(2013){Niu}, {Fu}, \& {Zong}}]{Niu2013}
{Niu}, J.-S., {Fu}, J.-N., \& {Zong}, W.-K. 2013, \raa, 13, 1181

\bibitem[{{Niu} {et~al.}(2023){Niu}, {Liu}, \& {Xue}}]{Niu2023}
{Niu}, J.-S., {Liu}, Y., \& {Xue}, H.-F. 2023, \aj, 166, 43

\bibitem[{{Niu} \& {Xue}(2022)}]{Niu2022}
{Niu}, J.-S. \& {Xue}, H.-F. 2022, \apjl, 938, L20

\bibitem[{{Niu} \& {Xue}(2024)}]{Niu2024}
{Niu}, J.-S. \& {Xue}, H.-F. 2024, \aap, 628, L8

\bibitem[{{P{\'a}pics} {et~al.}(2017){P{\'a}pics}, {Tkachenko}, {Van Reeth},
  {Aerts}, {Moravveji}, {Van de Sande}, {De Smedt}, {Bloemen}, {Southworth},
  {Debosscher}, {Niemczura}, \& {Gameiro}}]{Papics2017}
{P{\'a}pics}, P.~I., {Tkachenko}, A., {Van Reeth}, T., {et~al.} 2017, \aap,
  598, A74

\bibitem[{{Pietrukowicz} {et~al.}(2013){Pietrukowicz}, {Dziembowski},
  {Mr{\'o}z}, {Soszy{\'n}ski}, {Udalski}, {Poleski}, {Szyma{\'n}ski}, {Kubiak},
  {Pietrzy{\'n}ski}, {Wyrzykowski}, {Ulaczyk}, {Koz{\l}owski}, \&
  {Skowron}}]{Pietrukowicz2013}
{Pietrukowicz}, P., {Dziembowski}, W.~A., {Mr{\'o}z}, P., {et~al.} 2013,
  \actaa, 63, 379

\bibitem[{{Poretti} {et~al.}(2011){Poretti}, {Rainer}, {Weiss}, {Bogn{\'a}r},
  {Moya}, {Niemczura}, {Su{\'a}rez}, {Auvergne}, {Baglin}, {Baudin},
  {Benk{\H{o}}}, {Debosscher}, {Garrido}, {Mantegazza}, \&
  {Papar{\'o}}}]{Poretti2011}
{Poretti}, E., {Rainer}, M., {Weiss}, W.~W., {et~al.} 2011, \aap, 528, A147

\bibitem[{{Rathour} {et~al.}(2021){Rathour}, {Smolec}, \&
  {Netzel}}]{Rathour2021}
{Rathour}, R.~S., {Smolec}, R., \& {Netzel}, H. 2021, \mnras, 505, 5412

\bibitem[{{Steindl} {et~al.}(2022){Steindl}, {Zwintz}, \&
  {Vorobyov}}]{Steindl2022}
{Steindl}, T., {Zwintz}, K., \& {Vorobyov}, E. 2022, Nature Communications, 13,
  5355

\bibitem[{{Uytterhoeven} {et~al.}(2011){Uytterhoeven}, {Moya},
  {Grigahc{\`e}ne}, {Guzik}, {Guti{\'e}rrez-Soto}, {Smalley}, {Hand ler},
  {Balona}, {Niemczura}, {Fox Machado}, {Benatti}, {Chapellier}, {Tkachenko},
  {Szab{\'o}}, {Su{\'a}rez}, {Ripepi}, {Pascual}, {Mathias},
  {Mart{\'\i}n-Ru{\'\i}z}, {Lehmann}, {Jackiewicz}, {Hekker}, {Gruberbauer},
  {Garc{\'\i}a}, {Dumusque}, {D{\'\i}az-Fraile}, {Bradley}, {Antoci}, {Roth},
  {Leroy}, {Murphy}, {De Cat}, {Cuypers}, {Kjeldsen}, {Christensen-Dalsgaard},
  {Breger}, {Pigulski}, {Kiss}, {Still}, {Thompson}, \& {van
  Cleve}}]{Uytterhoeven2011}
{Uytterhoeven}, K., {Moya}, A., {Grigahc{\`e}ne}, A., {et~al.} 2011, \aap, 534,
  A125

\bibitem[{{Wils} {et~al.}(2008){Wils}, {Rozakis}, {Kleidis}, {Hambsch}, \&
  {Bernhard}}]{Wils2008}
{Wils}, P., {Rozakis}, I., {Kleidis}, S., {Hambsch}, F.~J., \& {Bernhard}, K.
  2008, \aap, 478, 865

\bibitem[{{Wu}(2001)}]{Wu2001}
{Wu}, Y. 2001, \mnras, 323, 248

\bibitem[{{Xue} {et~al.}(2018){Xue}, {Fu}, {Fox-Machado}, {Shi}, {Zhou},
  {Zhang}, {Michel}, {Yan}, {Niu}, {Zong}, {Su}, {Castro}, {Ayala-Loera}, \&
  {Altamirano-D{\'e}vora}}]{Xue2018}
{Xue}, H.-F., {Fu}, J.-N., {Fox-Machado}, L., {et~al.} 2018, \apj, 861, 96

\bibitem[{{Xue} \& {Niu}(2020)}]{Xue2020}
{Xue}, H.-F. \& {Niu}, J.-S. 2020, \apj, 904, 5

\bibitem[{{Xue} {et~al.}(2022){Xue}, {Niu}, \& {Fu}}]{Xue2022}
{Xue}, H.-F., {Niu}, J.-S., \& {Fu}, J.-N. 2022, \raa, 22, 105006

\bibitem[{{Xue} {et~al.}(2023){Xue}, {Niu}, {Xue}, \& {Yin}}]{Xue2023}
{Xue}, W., {Niu}, J.-S., {Xue}, H.-F., \& {Yin}, S. 2023, \raa, 23, 075002

\end{thebibliography}

\clearpage
\setcounter{figure}{0}
\setcounter{table}{0}
\renewcommand{\thefigure}{A\arabic{figure}}
\renewcommand{\thetable}{A\arabic{table}}
\onecolumn
\begin{appendix}

\section{Spectrograms}
\label{app:spectra}
\begin{figure*}[htp]
  \centering
  \includegraphics[width=0.48\textwidth]{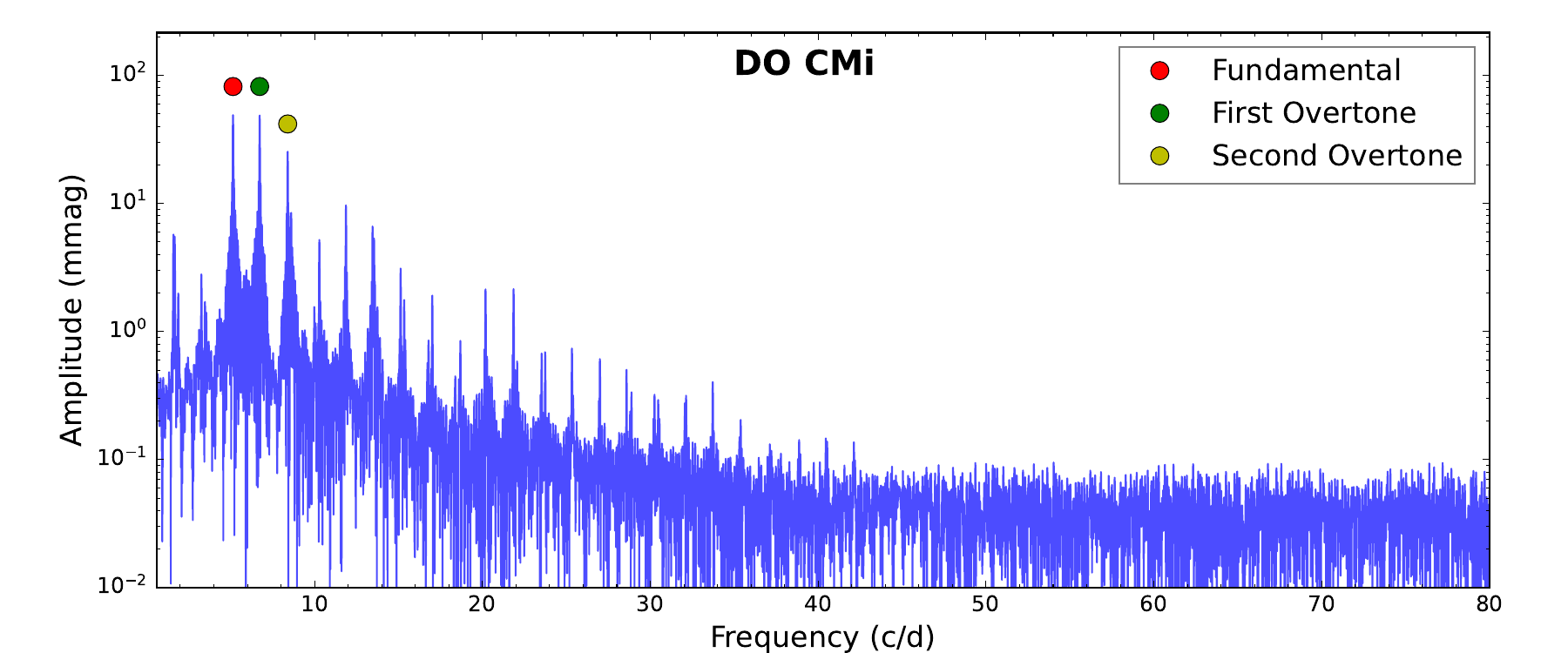}
  \includegraphics[width=0.48\textwidth]{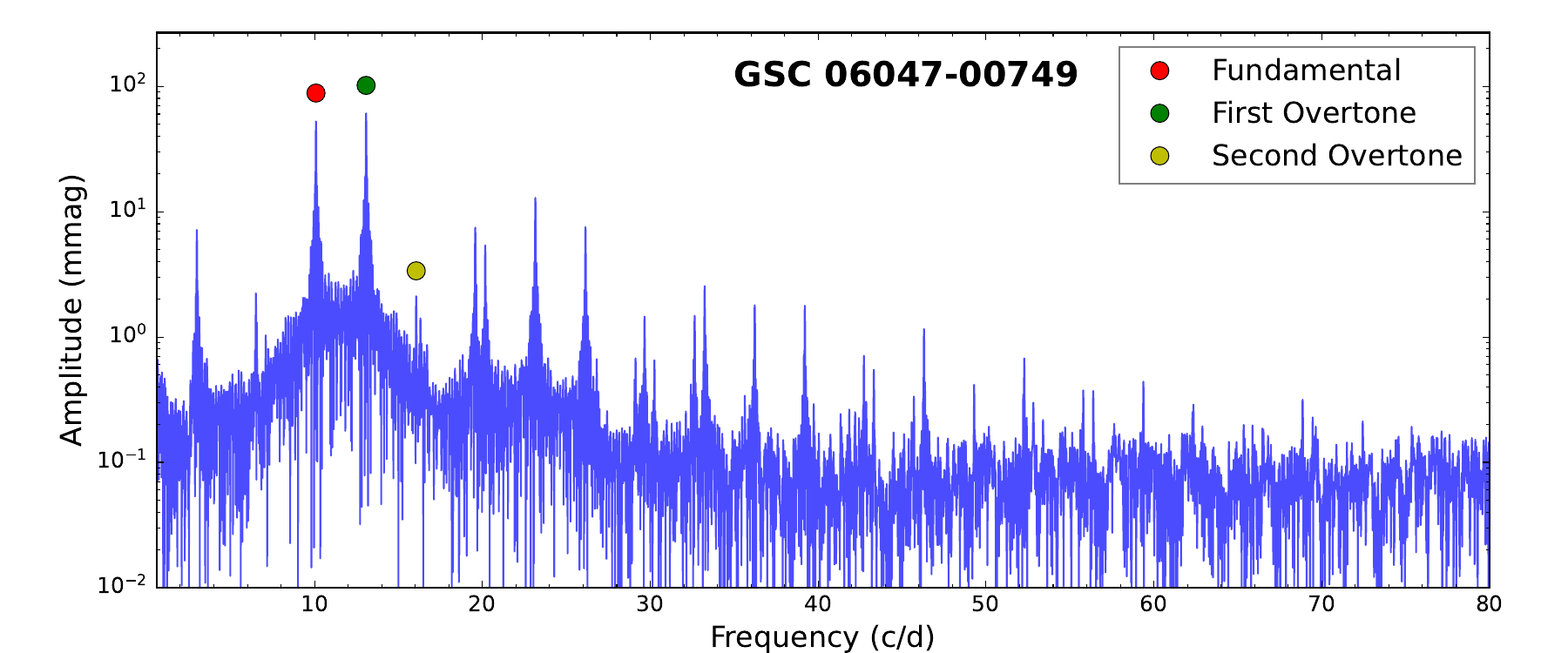}
  \includegraphics[width=0.48\textwidth]{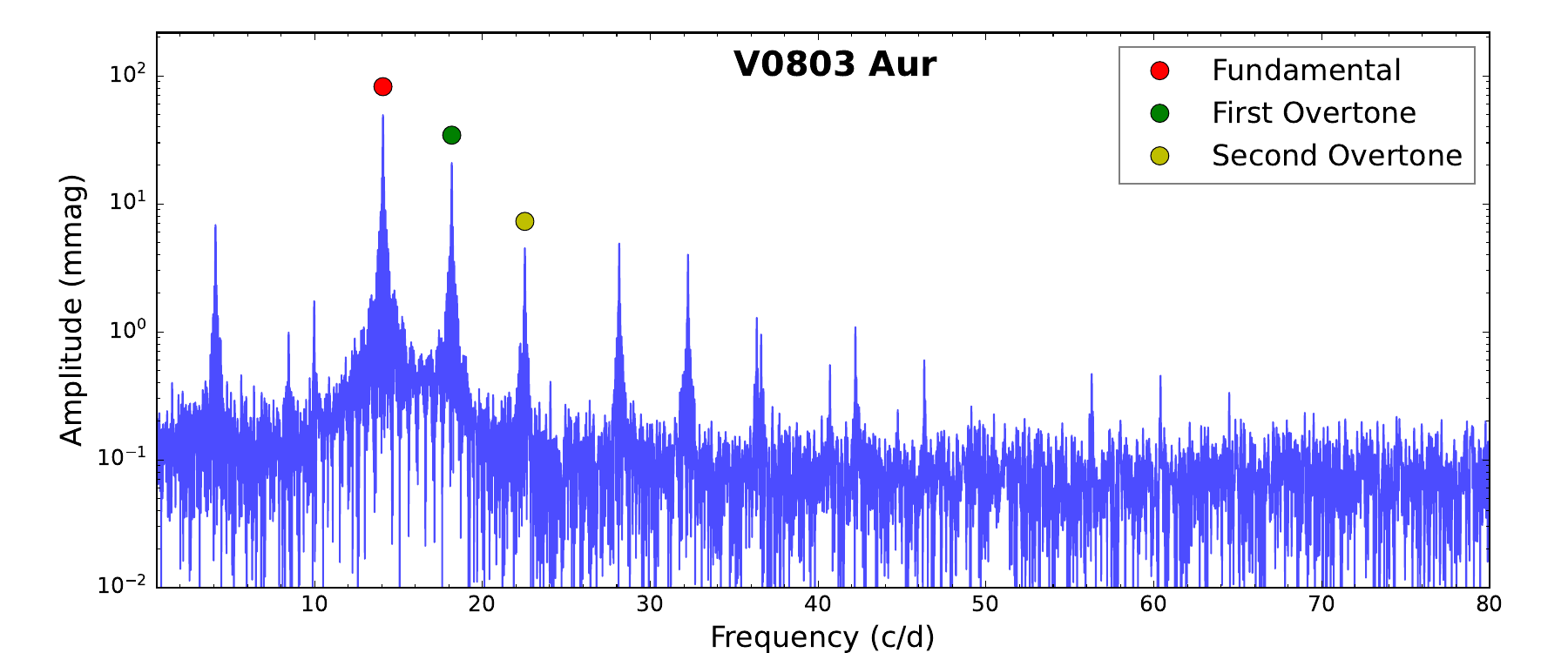}
  \includegraphics[width=0.48\textwidth]{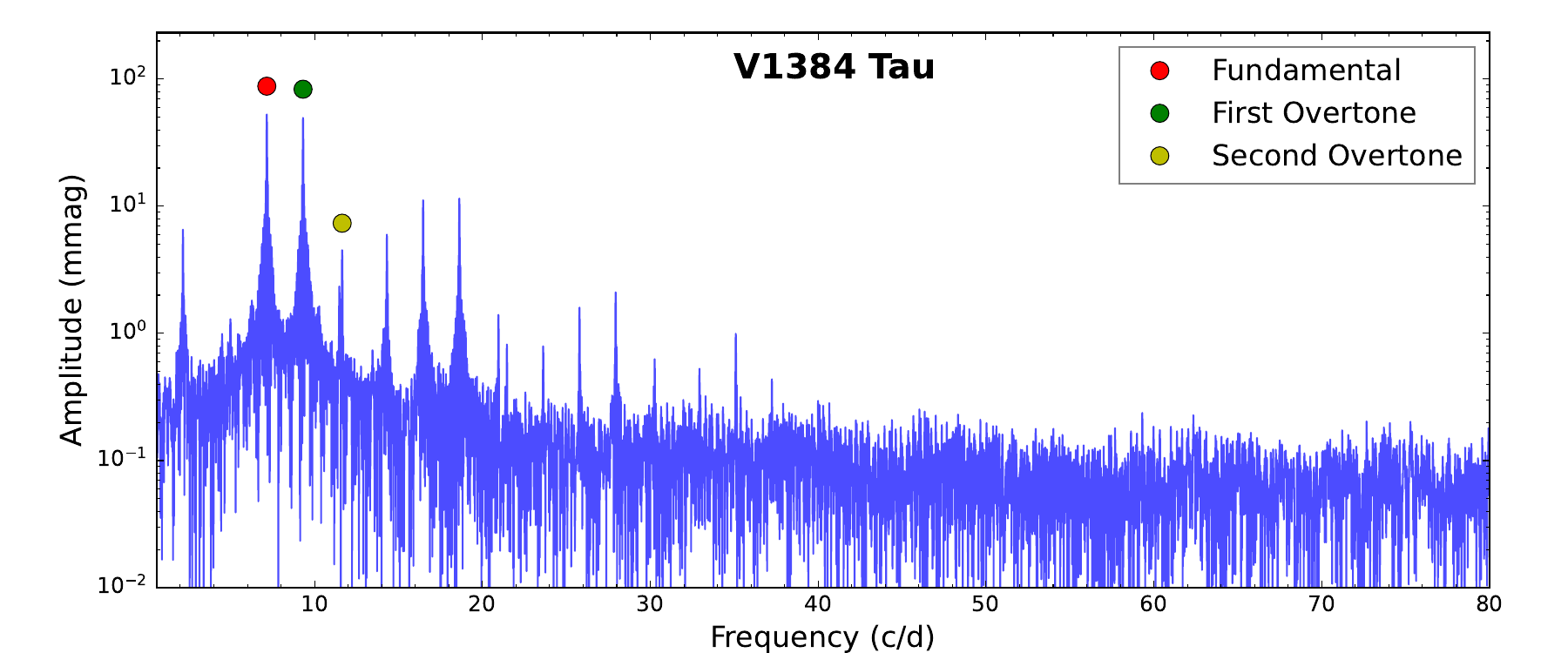}
  \includegraphics[width=0.48\textwidth]{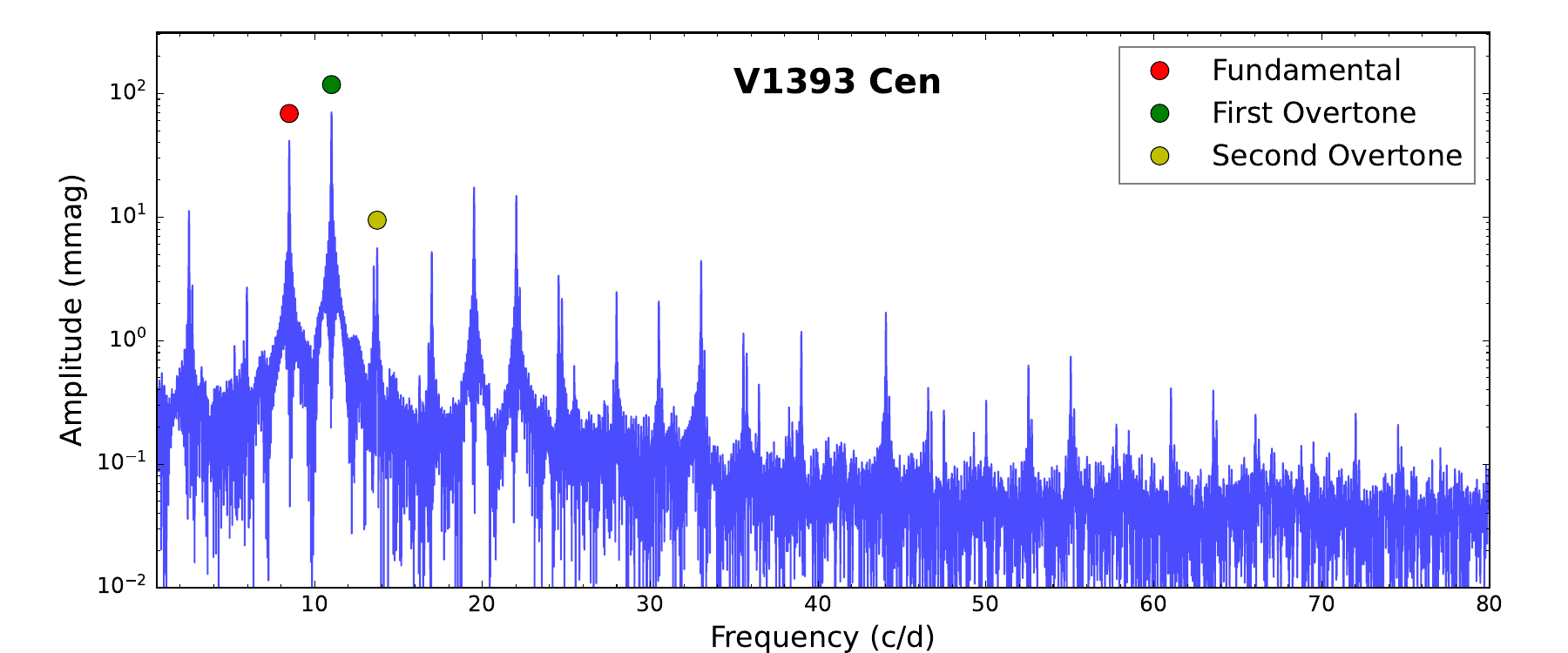}
  \caption{Spectrograms of the five HADS stars. The three independent frequencies for each of the them are marked as coloured dots.}
  \label{fig:spectra}
\end{figure*}

\section{Methods}
\label{app:method}

  All five HADS stars under investigation were observed by TESS, with their flux measurements obtained from the MAST Portal\footnote{https://mast.stsci.edu/portal/Mashup/Clients/Mast/Portal.html} and processed by the TESS Science Processing Operations Center (SPOC; \citet{Jenkins2016}). We converted the normalized fluxes to magnitudes using the TESS magnitude system and eliminated long-term trends for each sector. Subsequently, we selected the light curves with the most data points in a single sector for each star, as detailed in Table \ref{tab:basic_info}. {Fourier analysis was applied to these light curves, and the spectrograms of the five HADS stars are presented in Fig. \ref{fig:spectra}. Moreover, the resulting amplitudes and frequencies for the fundamental, first overtone, and second overtone pulsation modes ($A_0$, $A_1$, $A_2$ and $f_0$, $f_1$, $f_2$) are also presented in Table \ref{tab:basic_info}.}

\begin{figure*}[htp]
  \centering
  \includegraphics[width=0.48\textwidth]{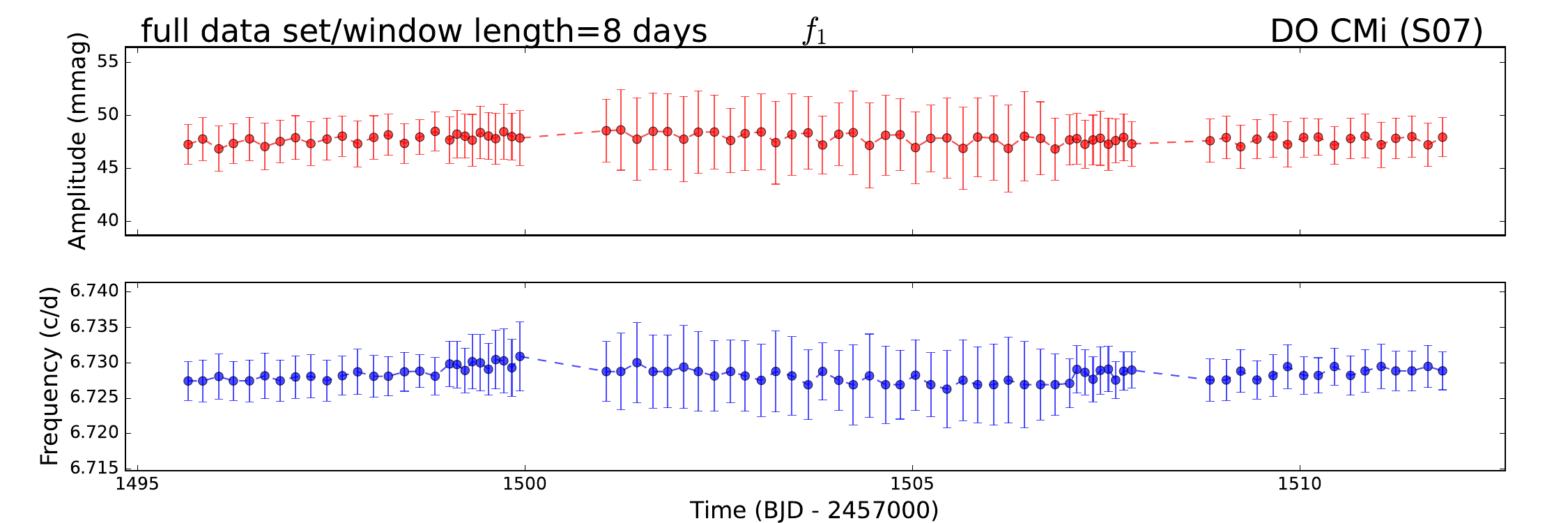}
  \includegraphics[width=0.48\textwidth]{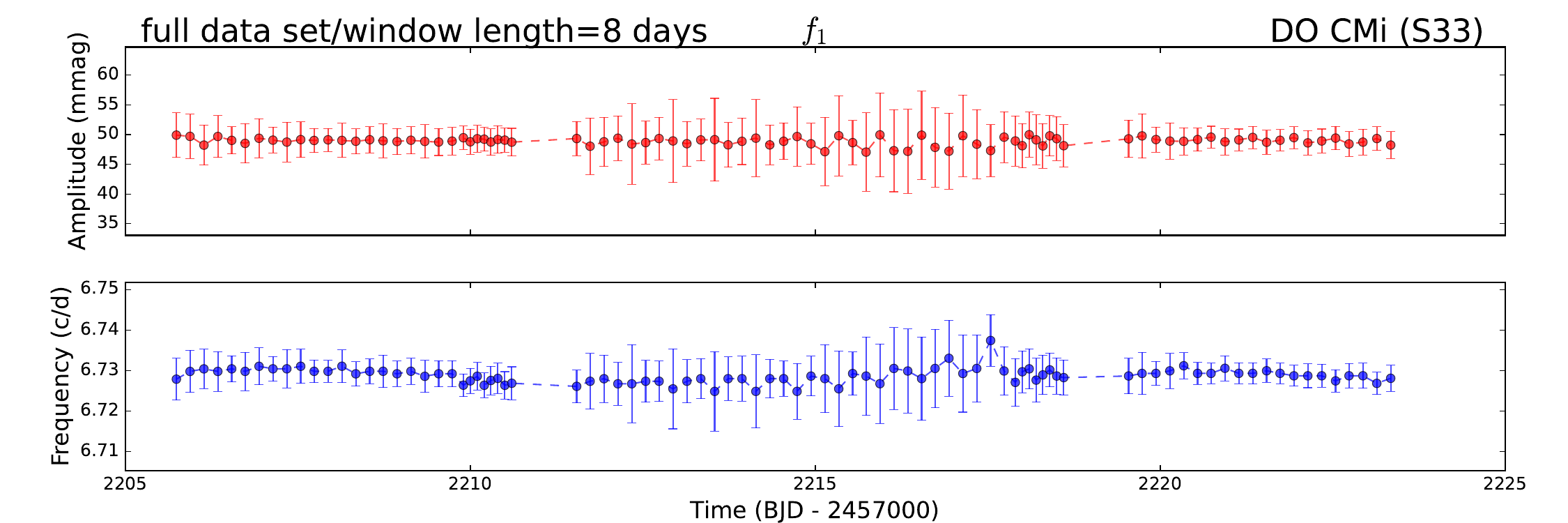}
  \includegraphics[width=0.48\textwidth]{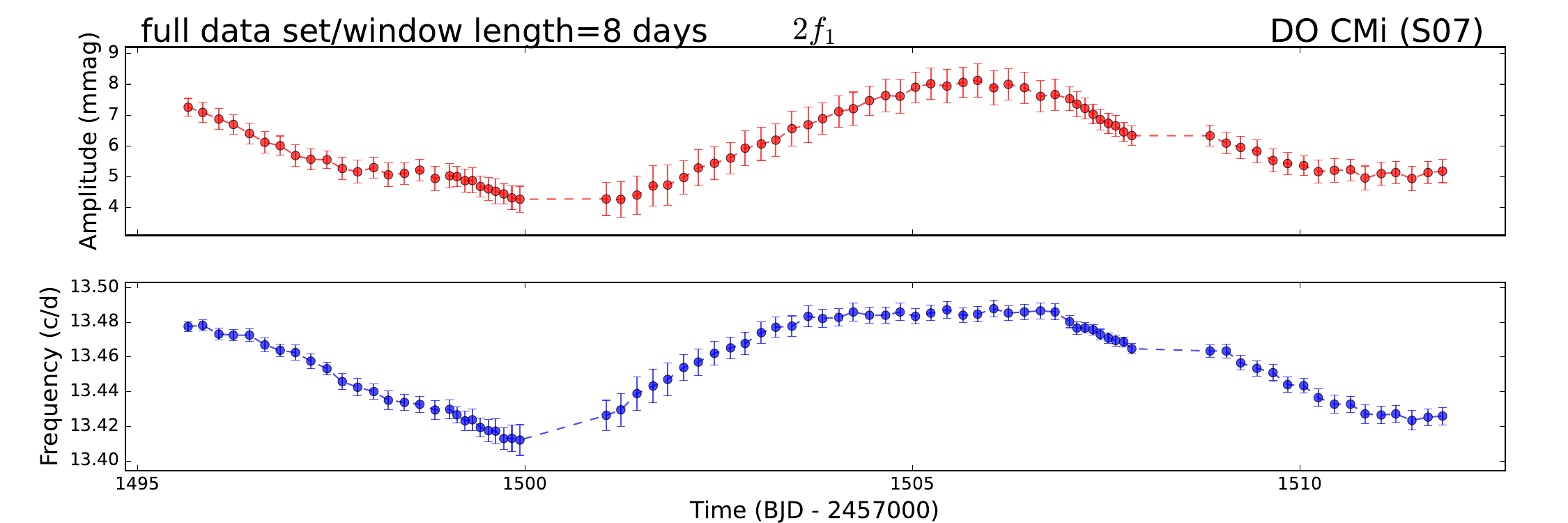}
  \includegraphics[width=0.48\textwidth]{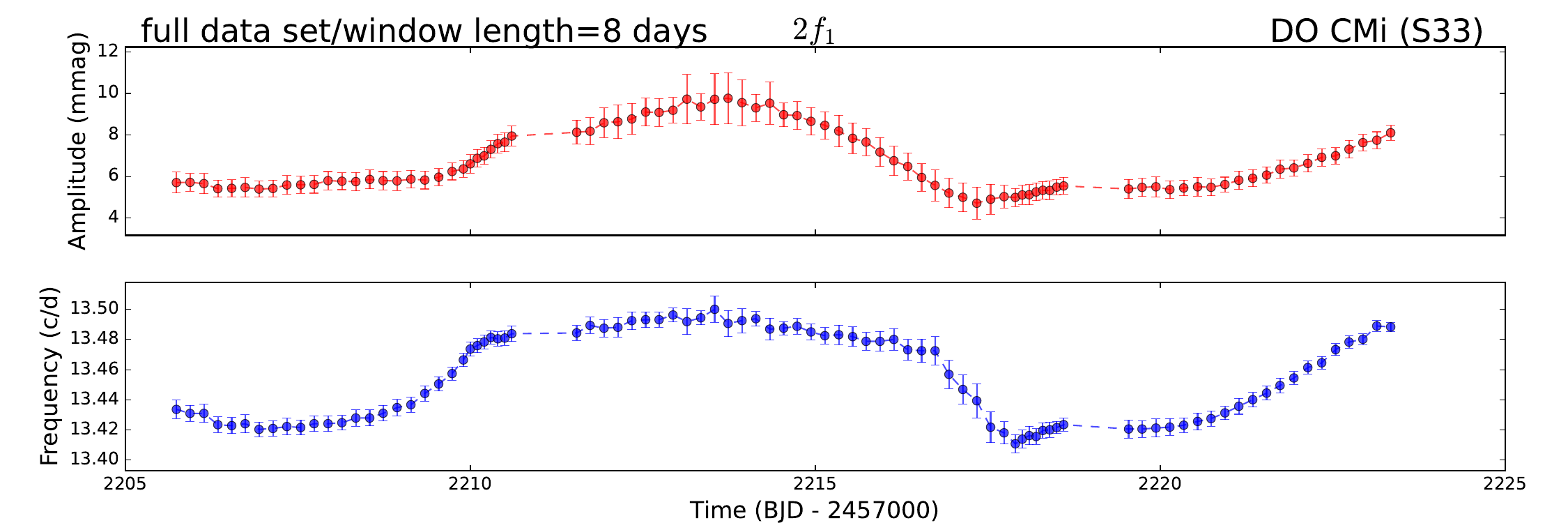}
  \includegraphics[width=0.48\textwidth]{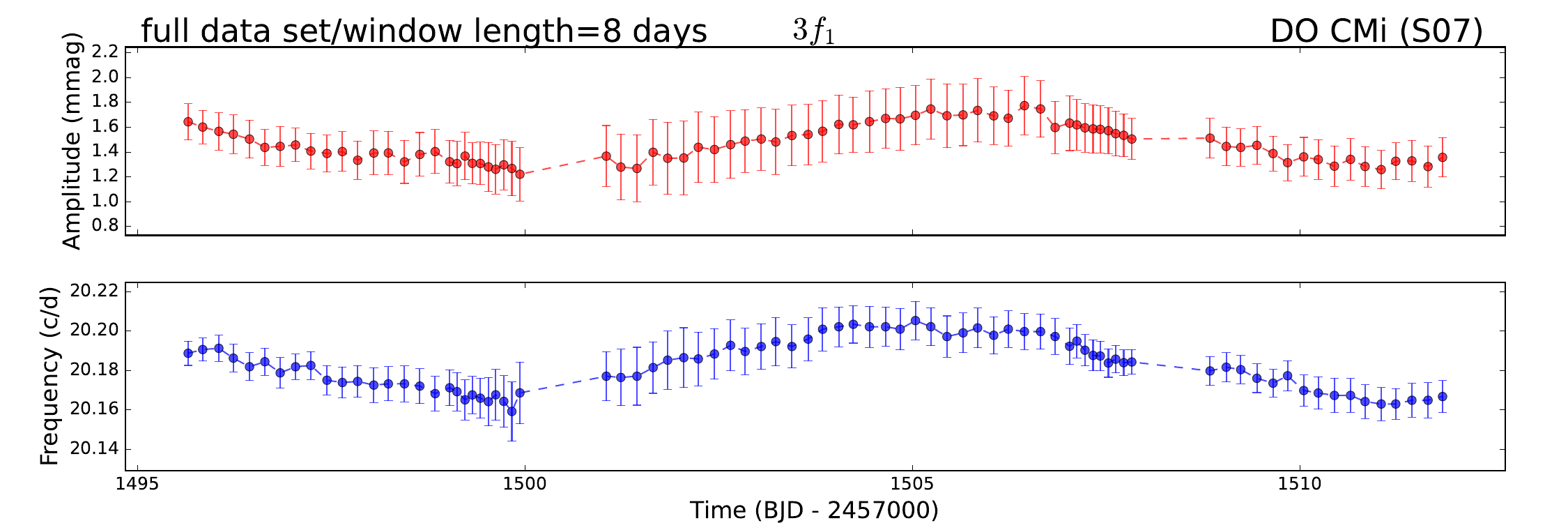}
  \includegraphics[width=0.48\textwidth]{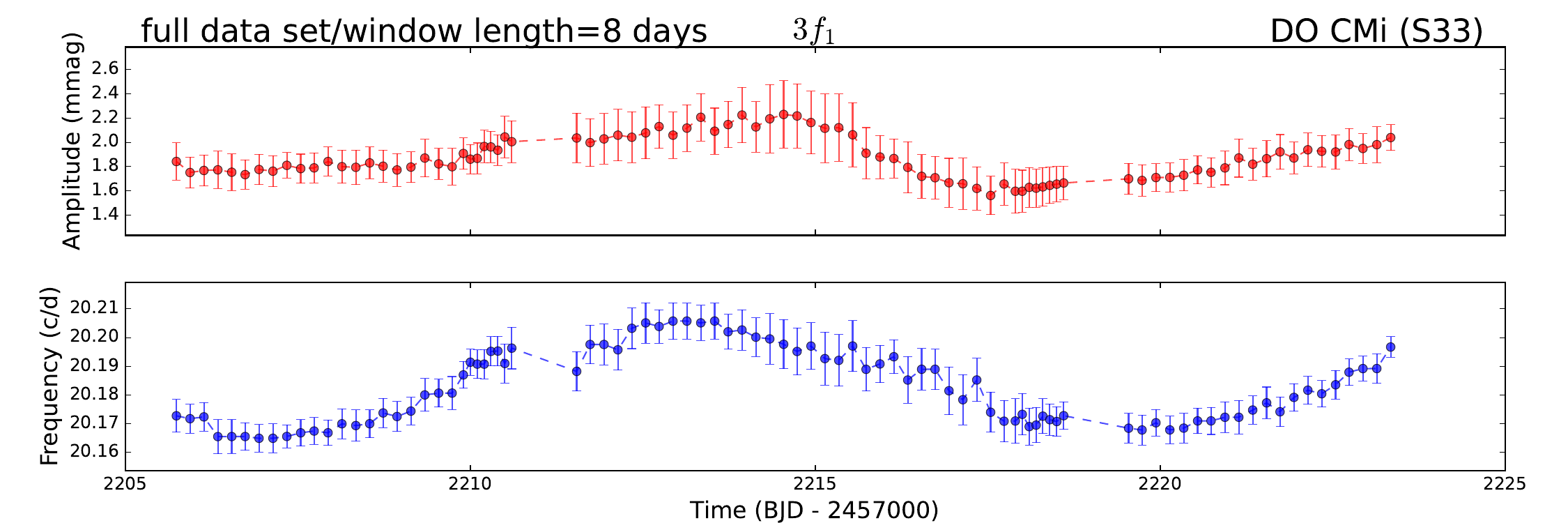}
  \caption{Amplitude and frequency variations in $f_1$ and its two harmonics for DO CMi, for TESS Sector 07 (left panel) and Sector 33 (right panel). We see that the detailed modulation phases in different sectors are different, but the correlation between the harmonics remains the same.}
  \label{fig:DOCMi_S07_S33}
\end{figure*}

\begin{figure*}[htp]
  \centering
  \includegraphics[width=0.48\textwidth]{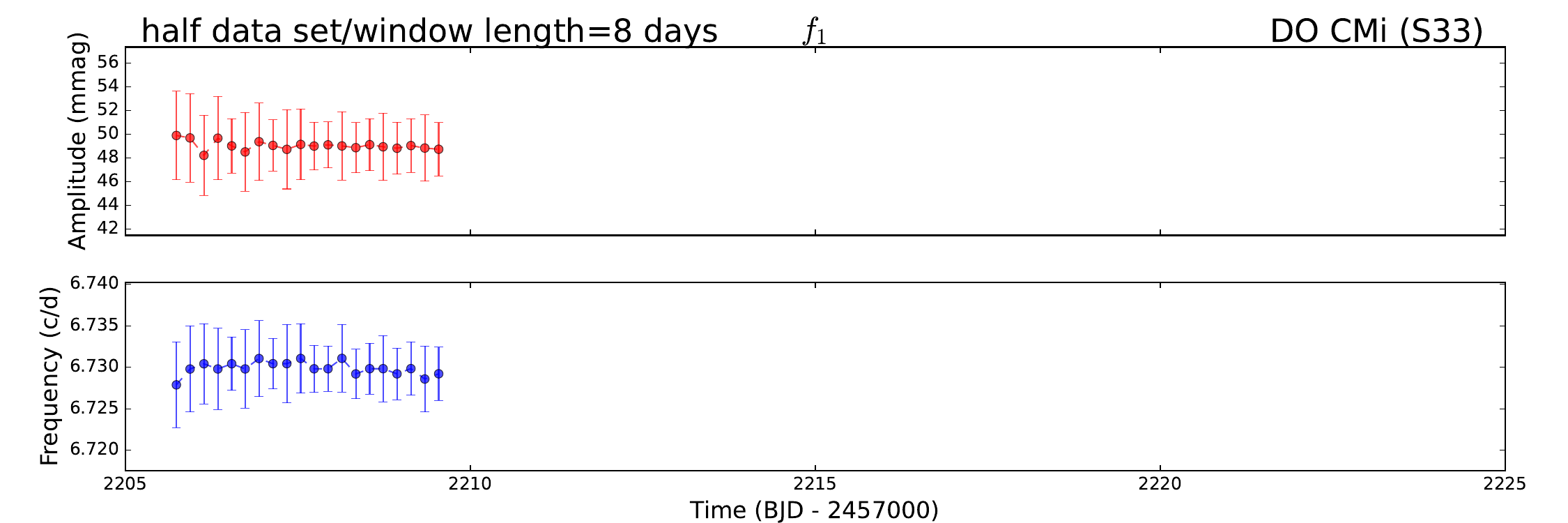}
  \includegraphics[width=0.48\textwidth]{fig/DOCMi_f1_s33_8d.pdf}
  \includegraphics[width=0.48\textwidth]{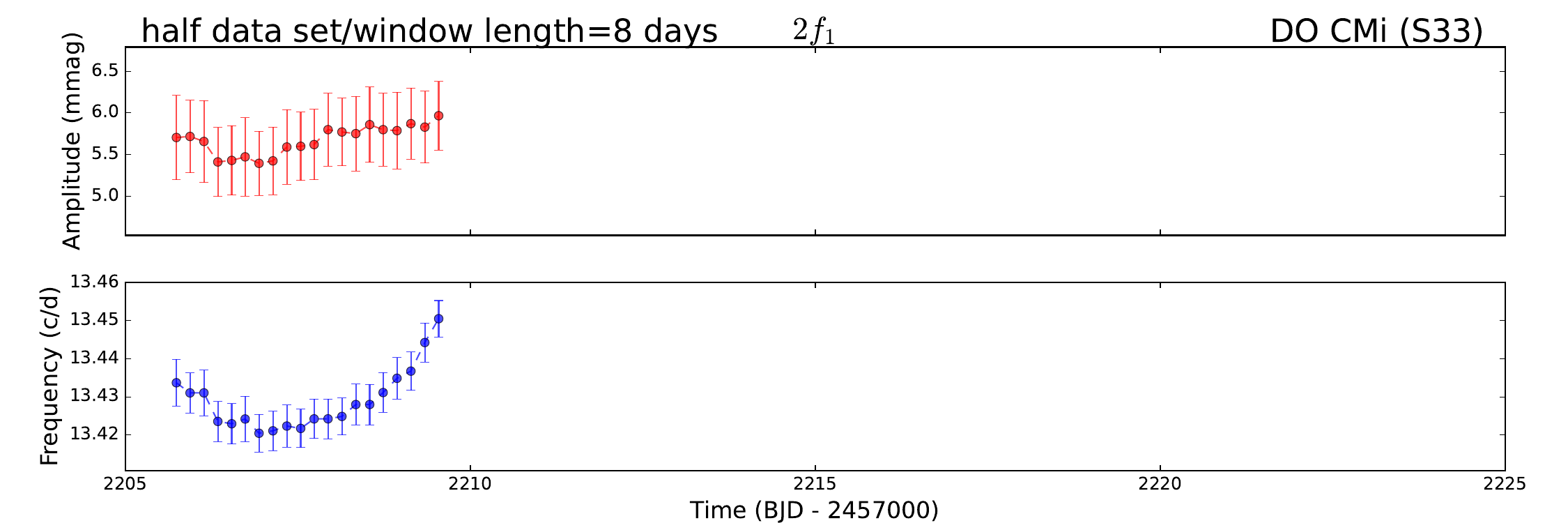}
  \includegraphics[width=0.48\textwidth]{fig/DOCMi_2f1_s33_8d.pdf}
  \includegraphics[width=0.48\textwidth]{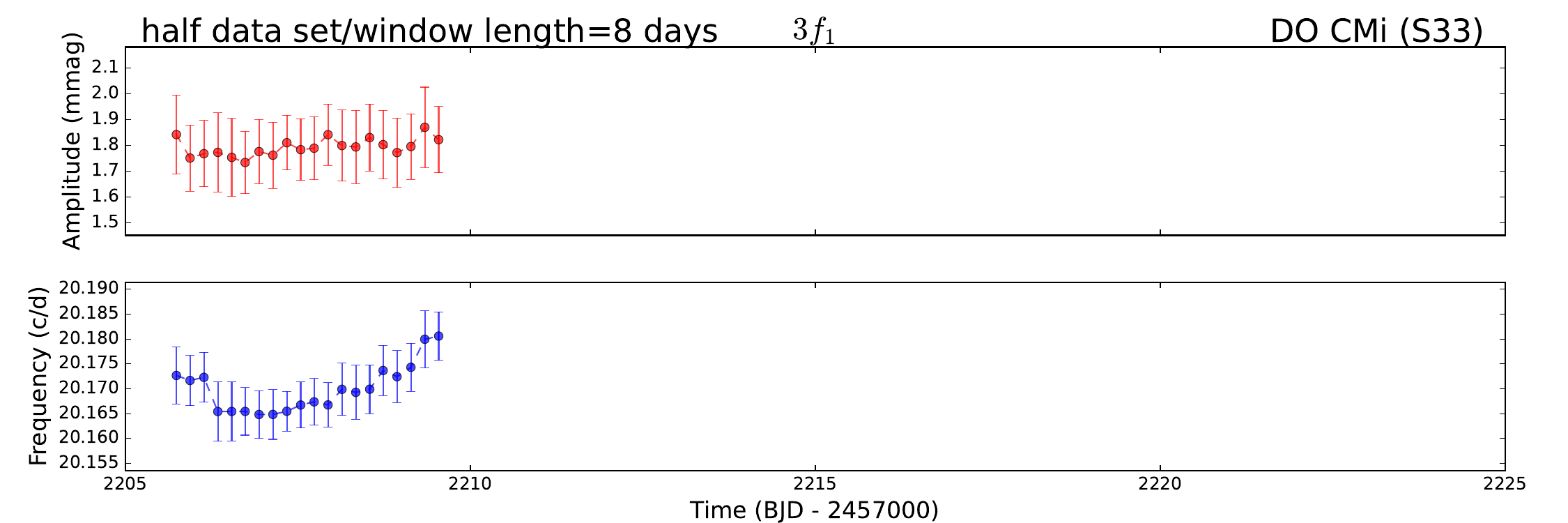}
  \includegraphics[width=0.48\textwidth]{fig/DOCMi_3f1_s33_8d.pdf}
  \caption{Amplitude and frequency variations in $f_1$ and its two harmonics for DO CMi, using half of the data points (left panel) and all of them (right panel) from Sector 33. We see that the modulation periods are not related to the data length.}
  \label{fig:DOCMi_S33_half}
\end{figure*}

\begin{figure*}[htp]
  \centering
  \includegraphics[width=0.33\textwidth]{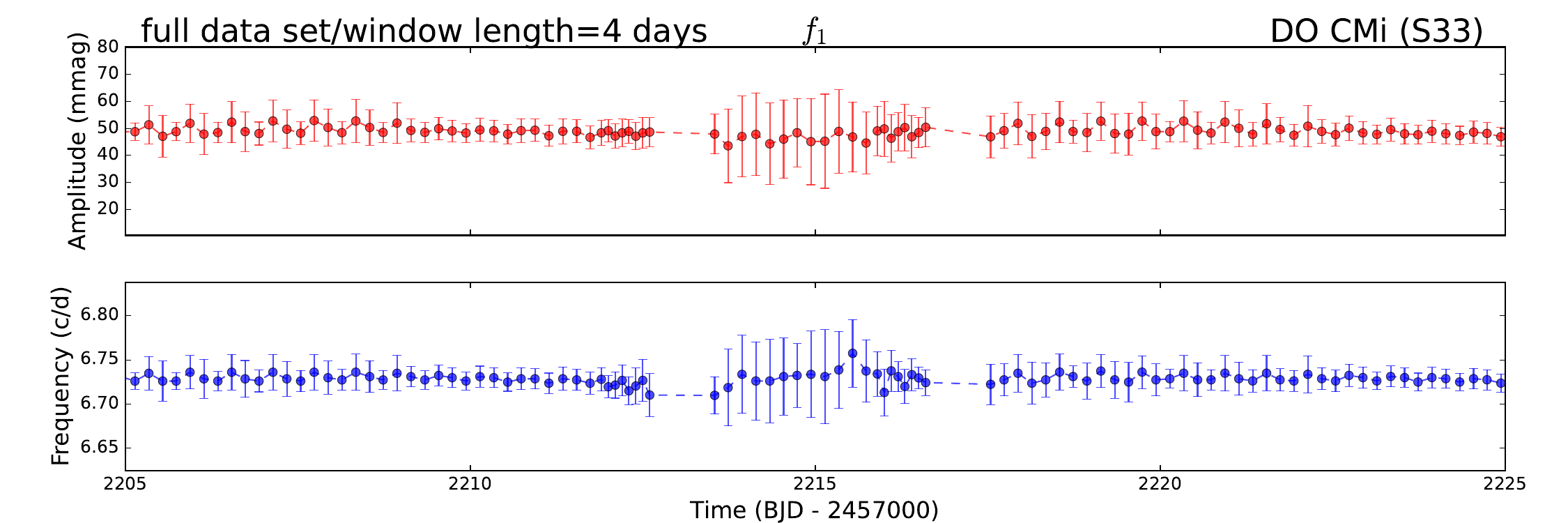}
  \includegraphics[width=0.33\textwidth]{fig/DOCMi_f1_s33_8d.pdf}
  \includegraphics[width=0.33\textwidth]{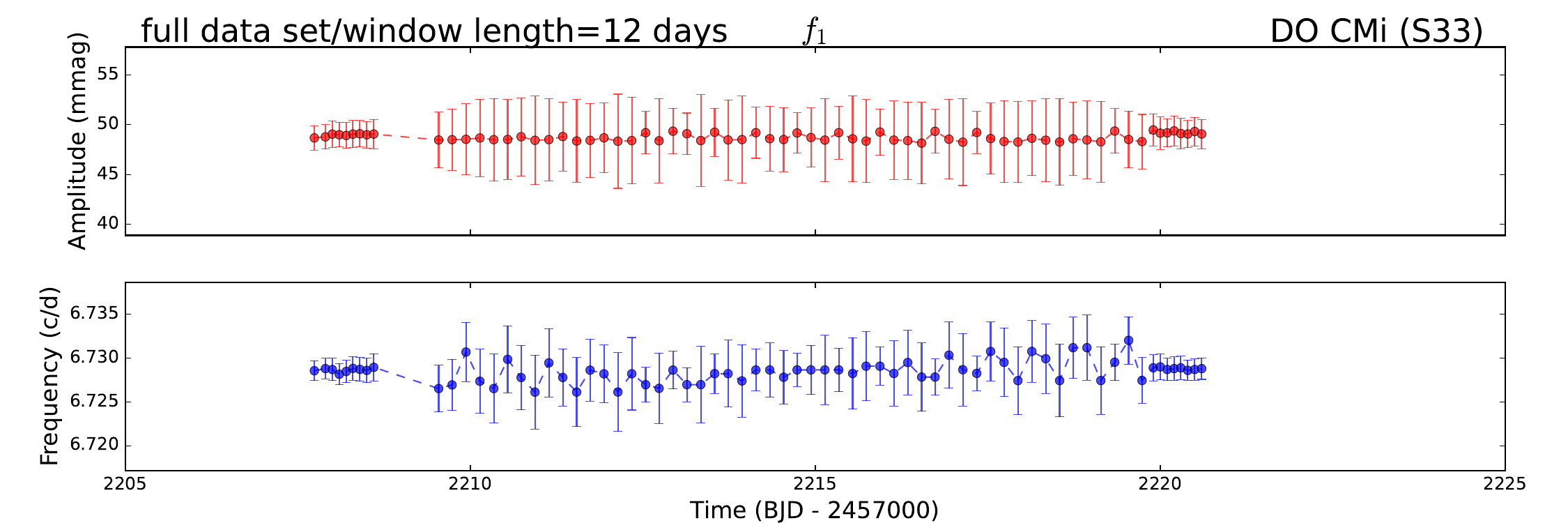}
  \includegraphics[width=0.33\textwidth]{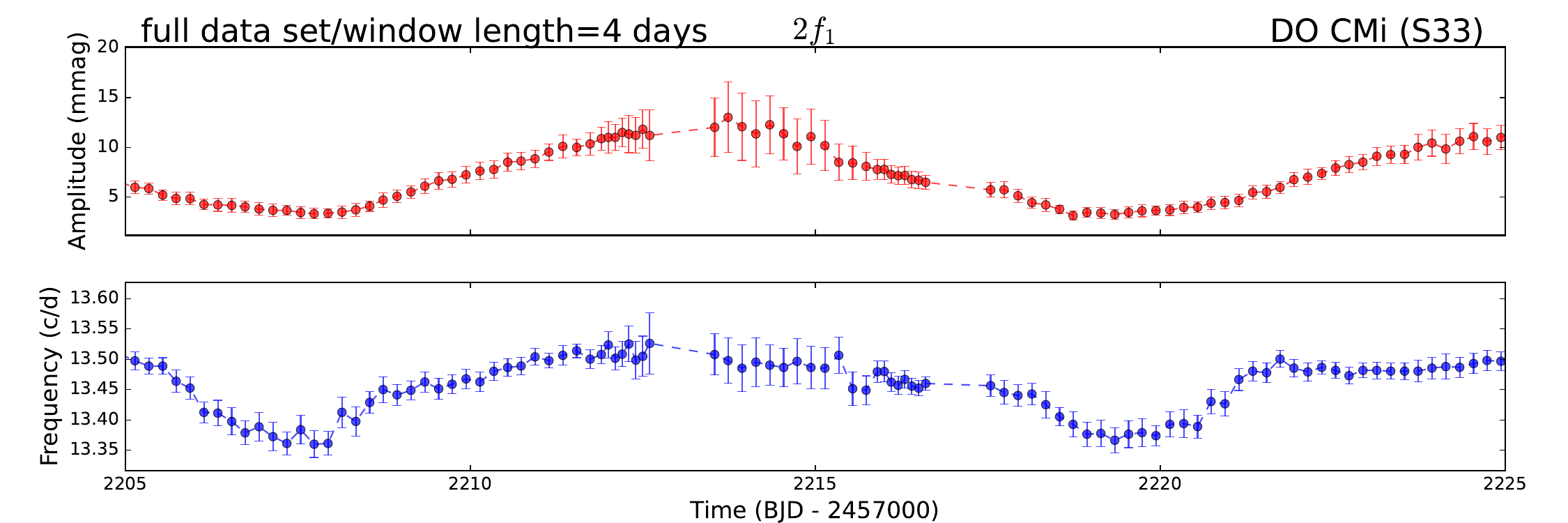}
  \includegraphics[width=0.33\textwidth]{fig/DOCMi_2f1_s33_8d.pdf}
  \includegraphics[width=0.33\textwidth]{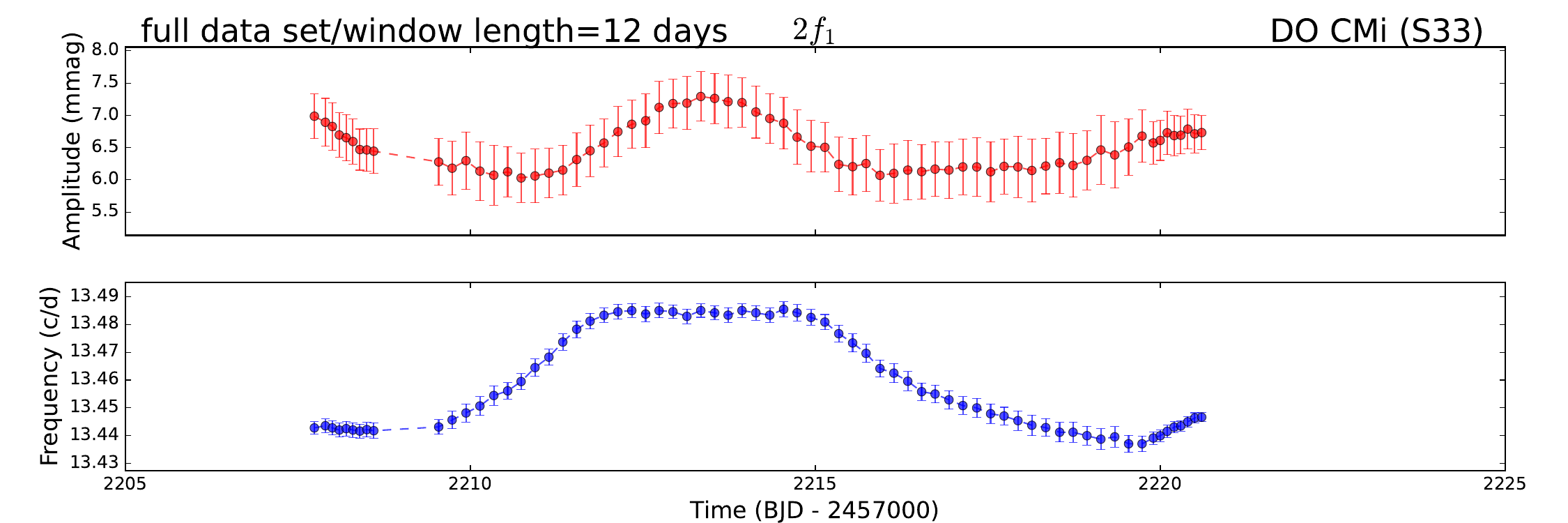}
  \includegraphics[width=0.33\textwidth]{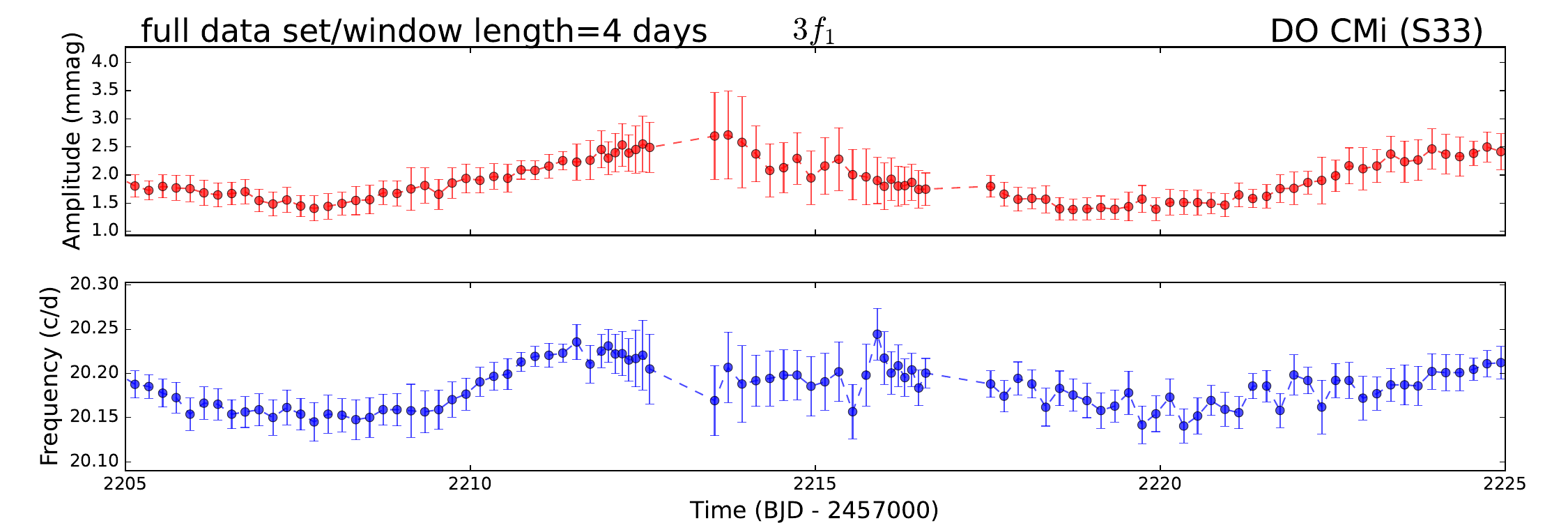}
  \includegraphics[width=0.33\textwidth]{fig/DOCMi_3f1_s33_8d.pdf}
  \includegraphics[width=0.33\textwidth]{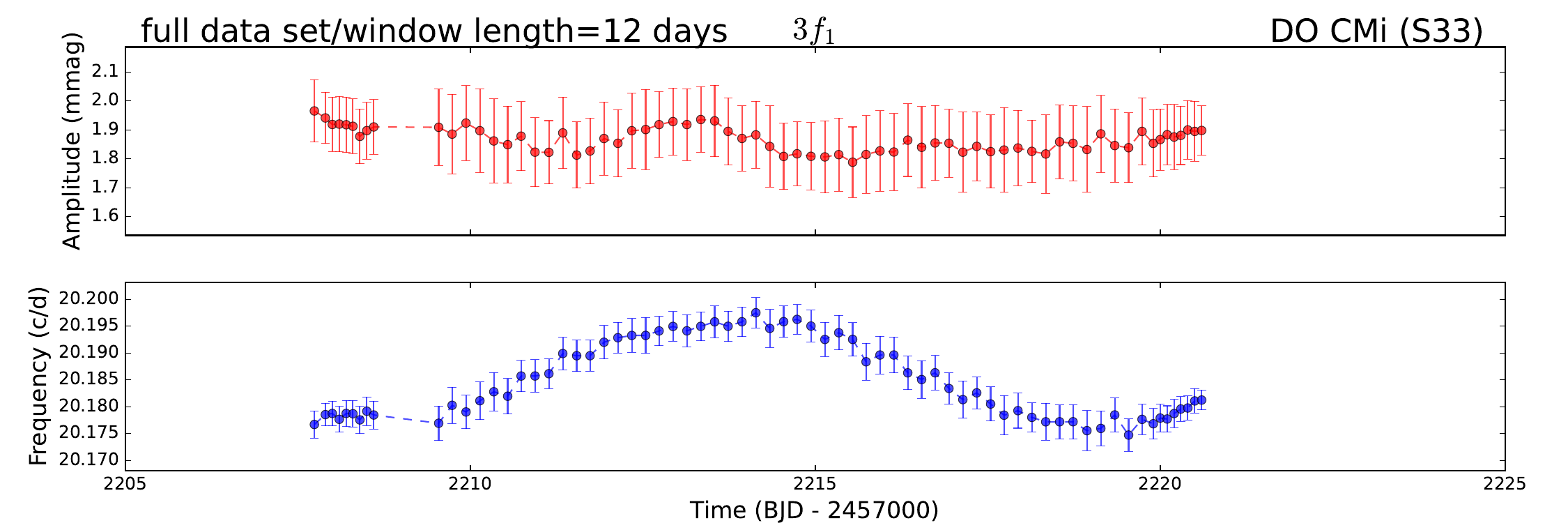}
  \caption{Amplitude and frequency variations in $f_1$ and its two harmonics for DO CMi, using different time window lengths: 4 days (left panel), 8 days (middle panel), and 12 days (right panel). We see that using different time windows can slightly affect the modulation phases and redistribute the lengths of the three segments of data points. However, the variation features are maintained.}
  \label{fig:DOCMi_4812}
\end{figure*}

\begin{figure*}[htp]
  \centering
  \includegraphics[width=0.5\textwidth]{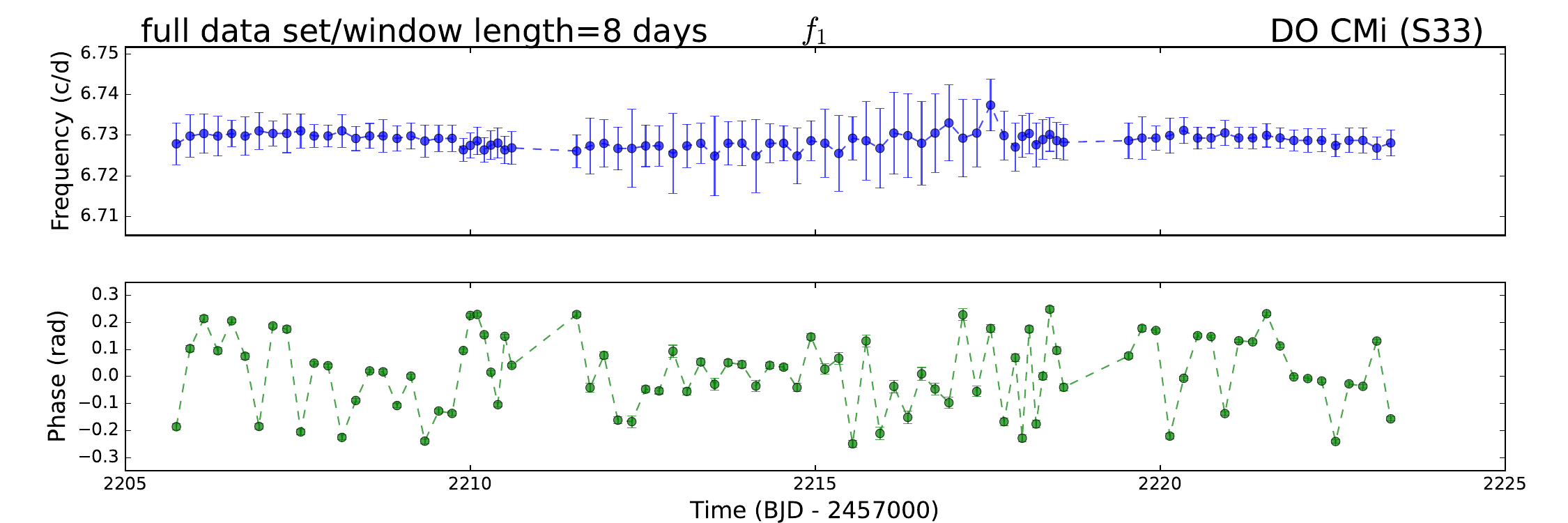}
  \includegraphics[width=0.5\textwidth]{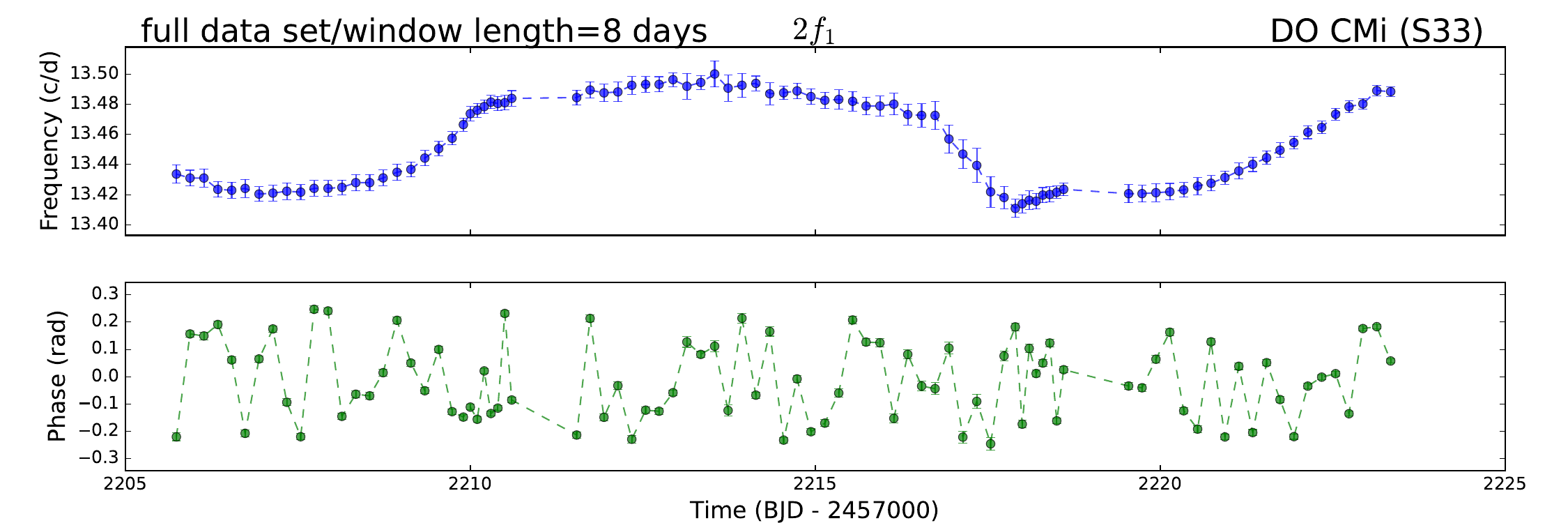}
  \includegraphics[width=0.5\textwidth]{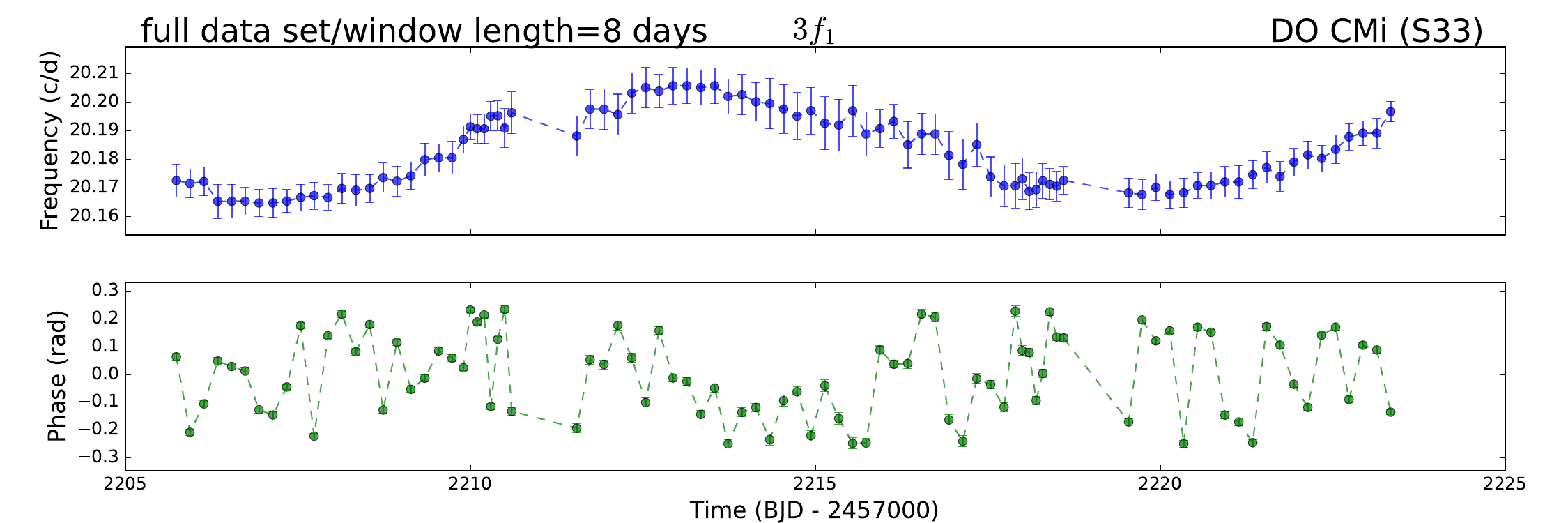}
  \caption{Frequency and phase variations in $f_1$ and its two harmonics for DO CMi. We see that the phases of the harmonics show violent and
  irregular fluctuations on short timescales (like the case in this study), and there are no clear correlations between these fluctuations from different harmonics.}
  \label{fig:DOCMi_phase}
\end{figure*}

  To elucidate the temporal variations in the amplitudes and frequencies of the harmonics, we employed the short-time Fourier transform, as described in detail by \citet{Niu2022,Niu2024} and \citet{Niu2023}. A pre-whitening process was executed within an 8-day time window, which was advanced in increments of 0.2 days across the entire dataset.\footnote{The choices of the time window length and dataset do not affect the conclusions of the present work, which are tested by $f_1$ and its harmonics of DO CMi in Figs. \ref{fig:DOCMi_S07_S33}, \ref{fig:DOCMi_S33_half}, and \ref{fig:DOCMi_4812}.} At each step, the pre-whitening process was utilized to obtain the amplitudes and frequencies of the independent pulsation modes and their harmonics, with the phase, not the primary focus of this study, left as a free parameter\footnote{The phases of the harmonics on short timescales show violent and irregular fluctuations, and there are no clear correlations between these fluctuations from different harmonics. See, e.g., Fig. \ref{fig:DOCMi_phase} for $f_1$ and its harmonics of DO CMi.}
    
  The pre-whitening process was performed through Fourier decomposition, as represented by the following equation:
    \begin{equation}
      \label{eq:Fourier_de}
      m = m_{0} + \sum A_{i} \sin \left[ 2 \pi (f_{i} t + \phi_{i}) \right],
    \end{equation}
  where $m_0$ is the shifted value, $A_{i}$  the amplitude, $f_{i}$  the frequency, and $\phi_{i}$  the phase.
  The amplitude uncertainties ($\sigma_{a}$) were determined as the median amplitude value within a spectral window of 4 $\cd$, divided equally by the frequency peak, in line with methods described by \citet{Niu2023} and \citet{Niu2024}. Frequency uncertainties ($\sigma_{f}$) were estimated based on the relationship between amplitude and frequency as proposed by \citet{Montgomery1999} and \citet{Aerts2021}, as shown in the equation below:
  \begin{equation}
      \label{eq:sigma_f}
      \sigma_{f} = \sigma_{a} \cdot \frac{\sqrt{3}}{\pi \cdot A \cdot T},
    \end{equation}
  where $A$ is the amplitude and $T$ represents the total time baseline used in the pre-whitening process.

\section{Amplitude and frequency variations}
\label{app:variations}

\begin{figure*}[htp]
  \centering
  \includegraphics[width=0.48\textwidth]{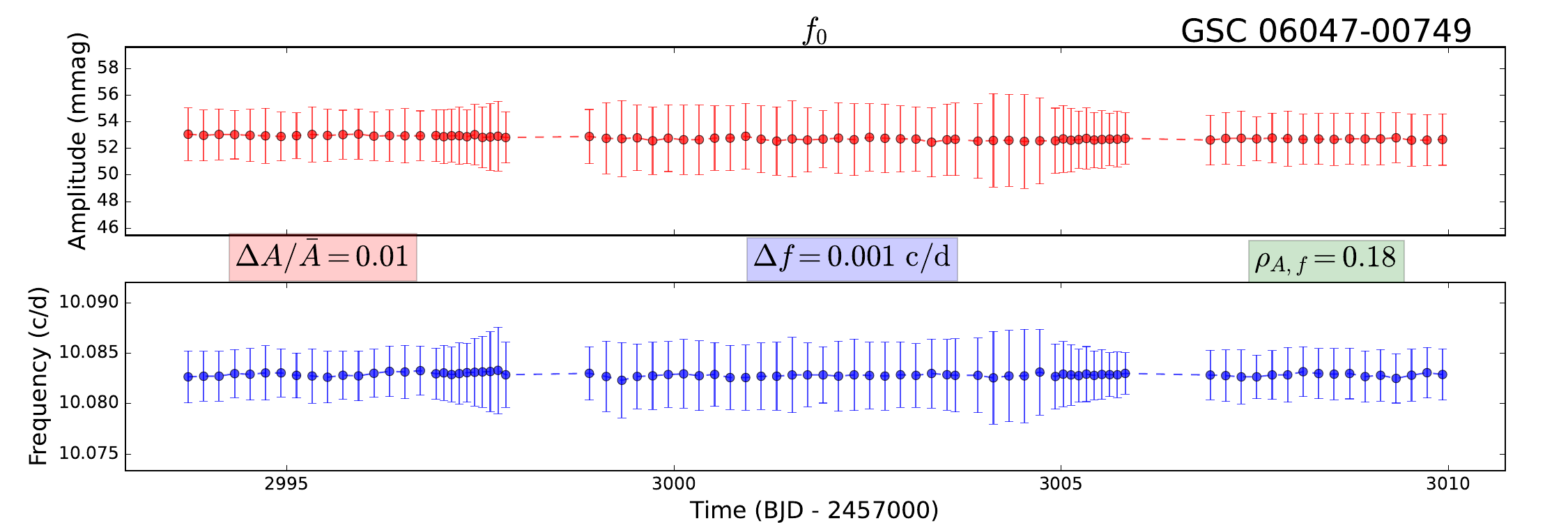}
  \includegraphics[width=0.48\textwidth]{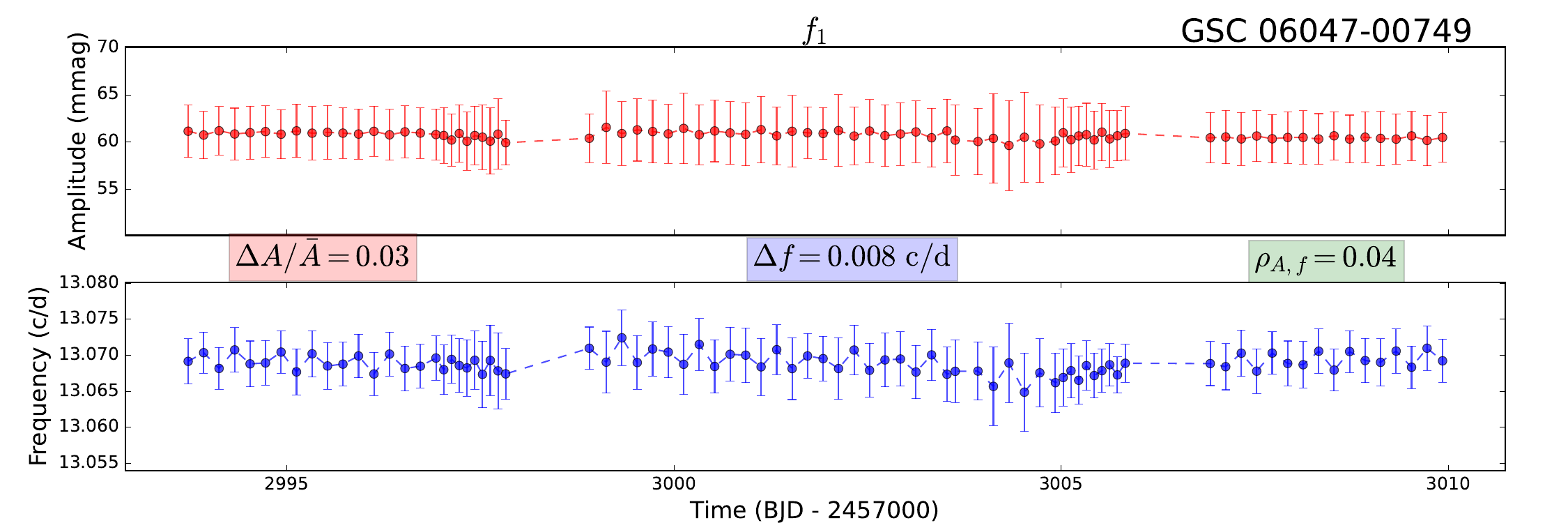}
  \includegraphics[width=0.48\textwidth]{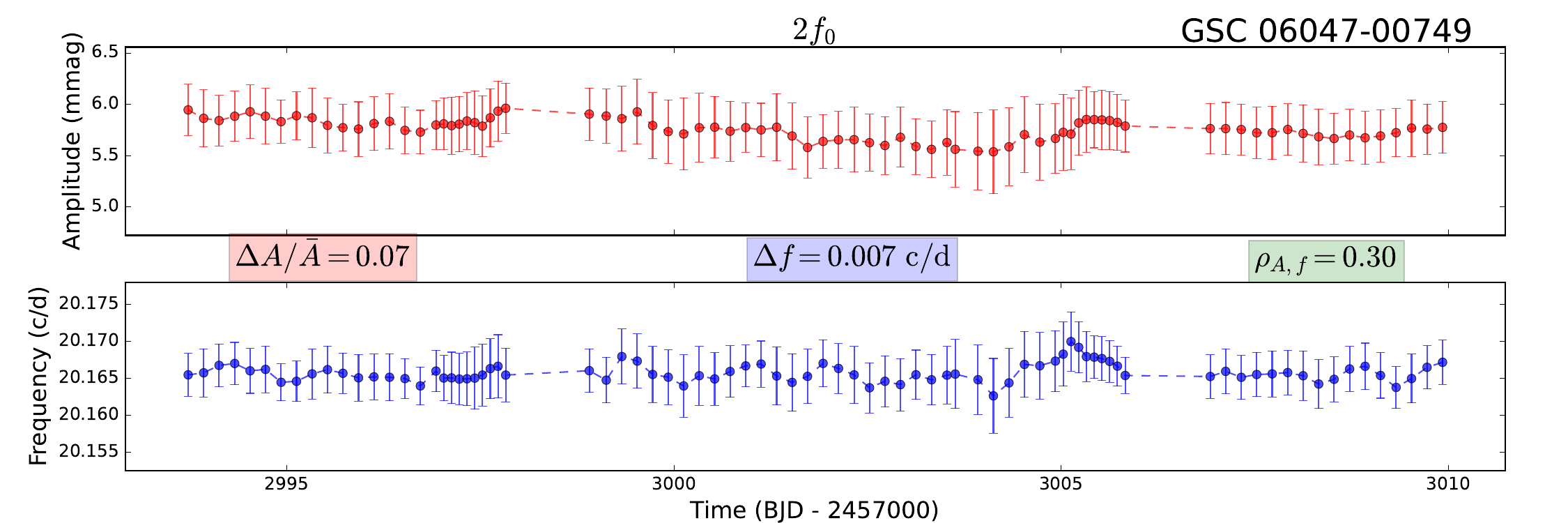}
  \includegraphics[width=0.48\textwidth]{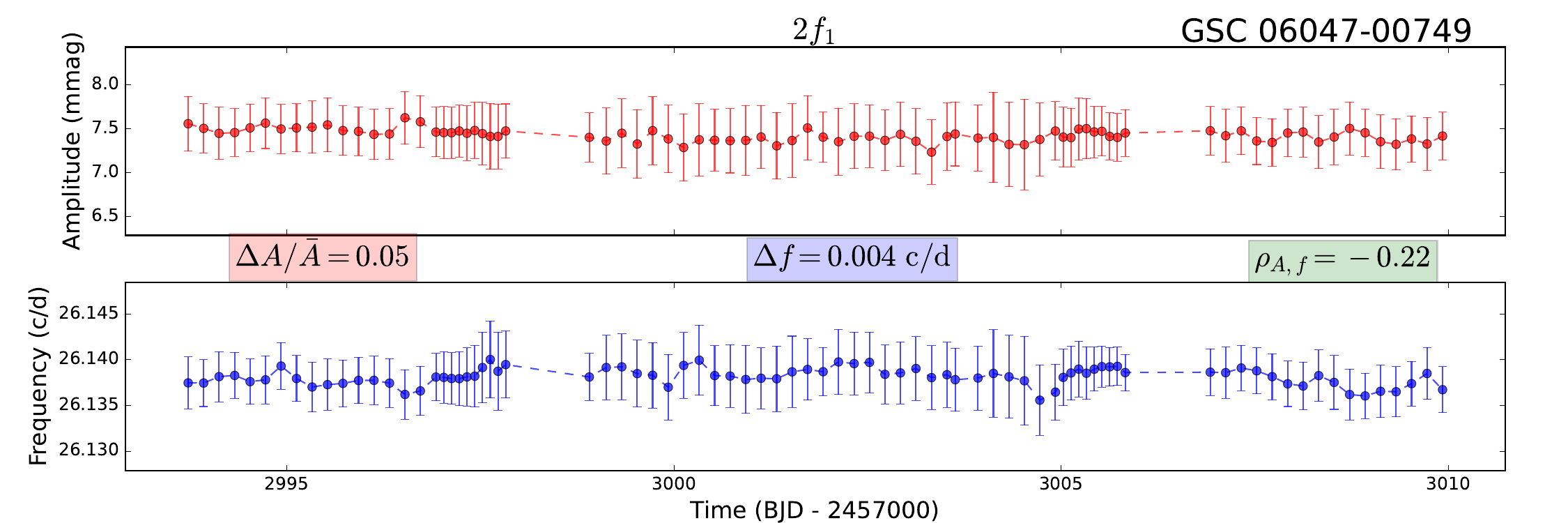}
  \includegraphics[width=0.48\textwidth]{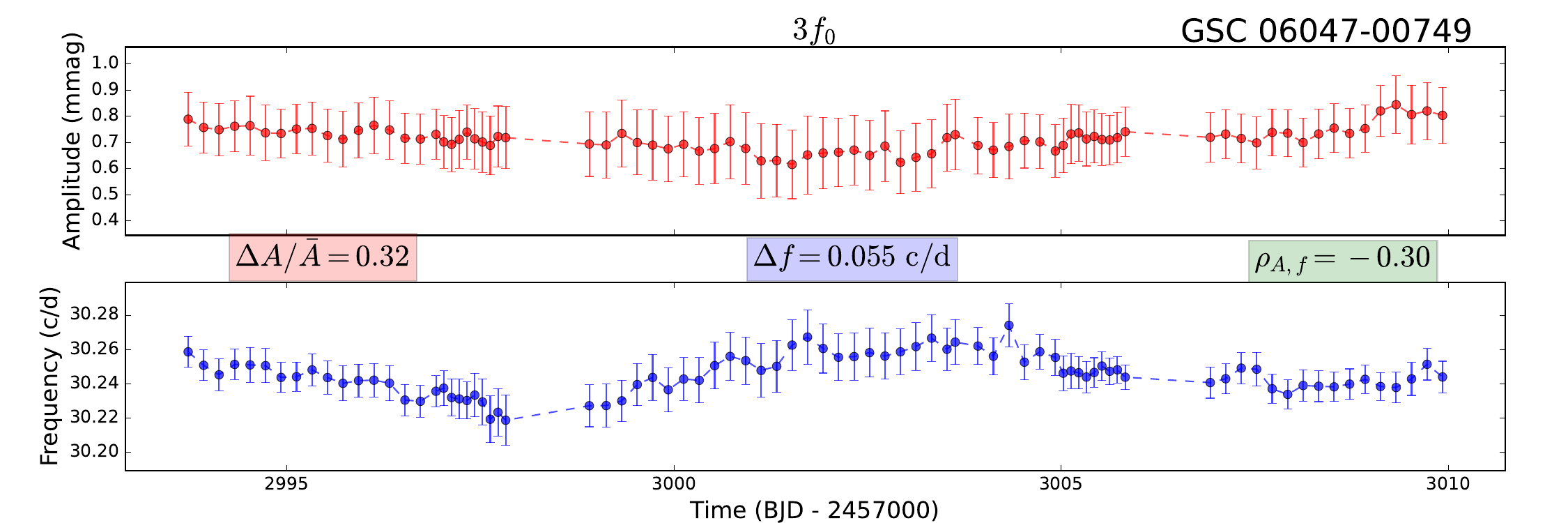}
  \includegraphics[width=0.48\textwidth]{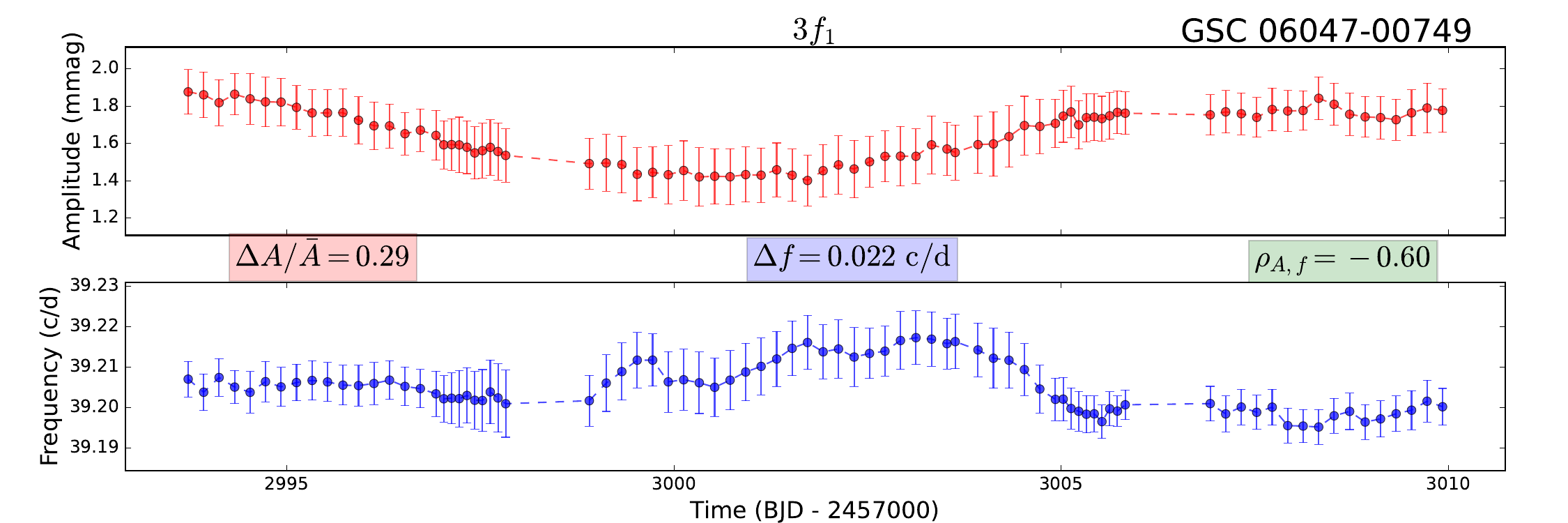}
  \phantom{\includegraphics[width=0.48\textwidth]{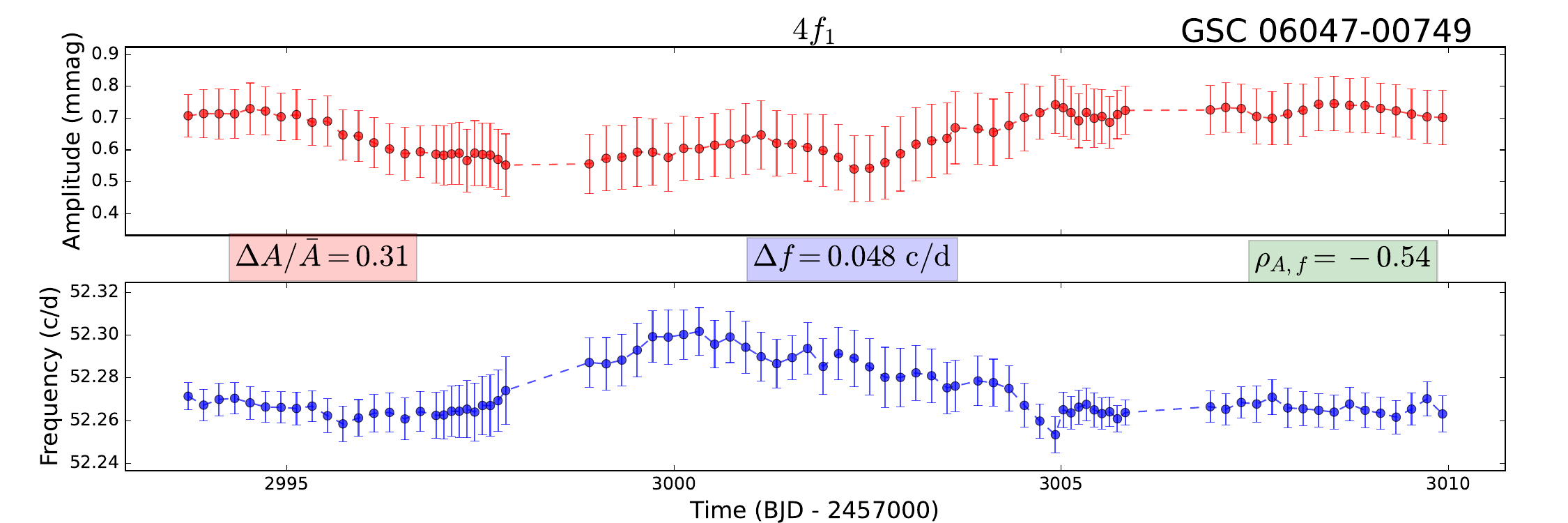}}
  \includegraphics[width=0.48\textwidth]{fig/GSC06047-00749_4f1.pdf}
  \caption{Amplitude and frequency variations in the fundamental and first overtones and their harmonic pulsation modes ({$f_{0}$ through $3f_{0}$} and {$f_{1}$ through $4f_{1}$}) in GSC 06047-00749.}
  \label{fig:GSC06047-00749}
\end{figure*}

\begin{figure*}[htp]
  \centering
  \includegraphics[width=0.48\textwidth]{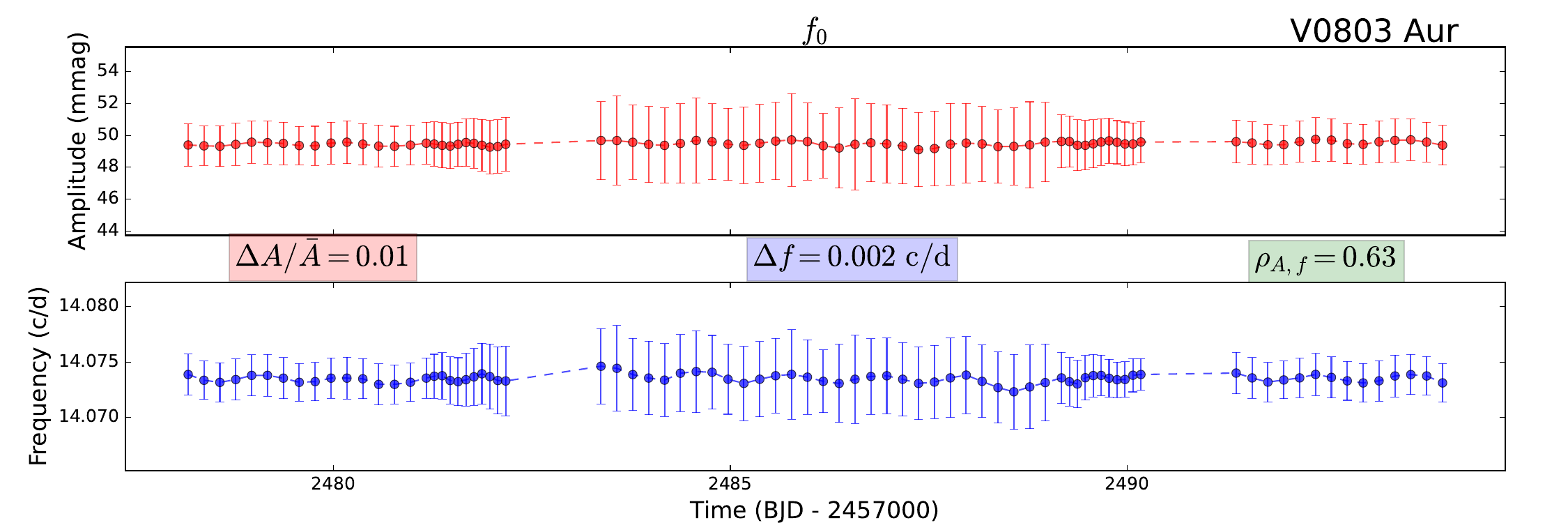}
  \includegraphics[width=0.48\textwidth]{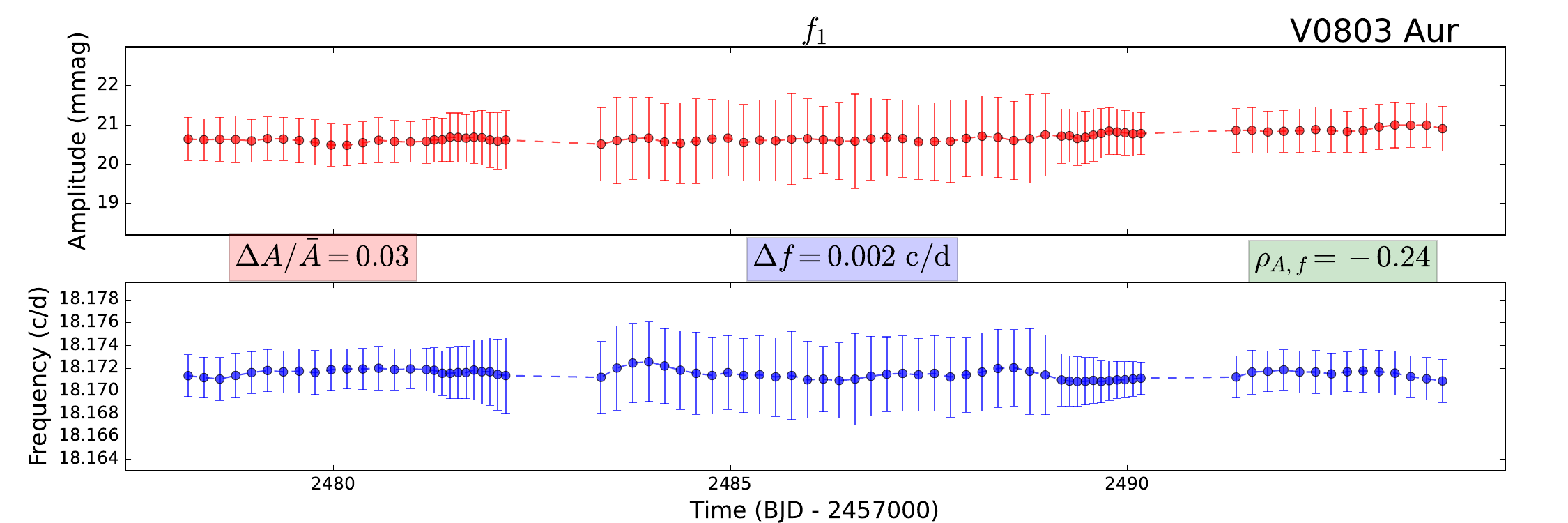}
  \includegraphics[width=0.48\textwidth]{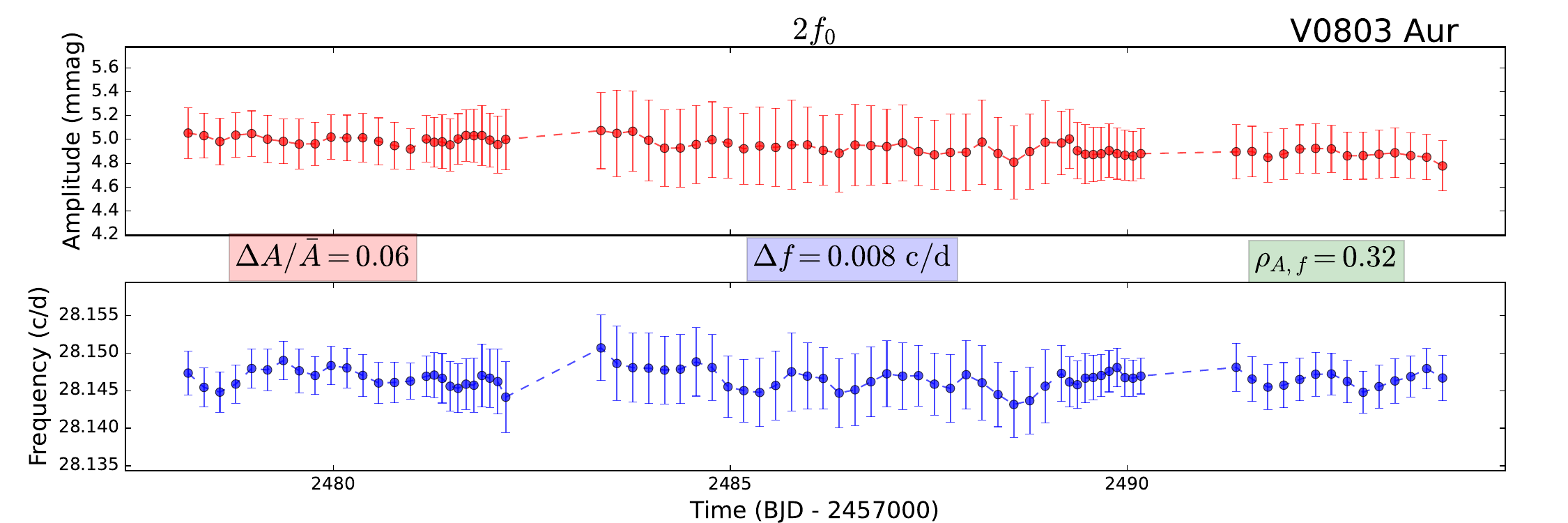}
  \includegraphics[width=0.48\textwidth]{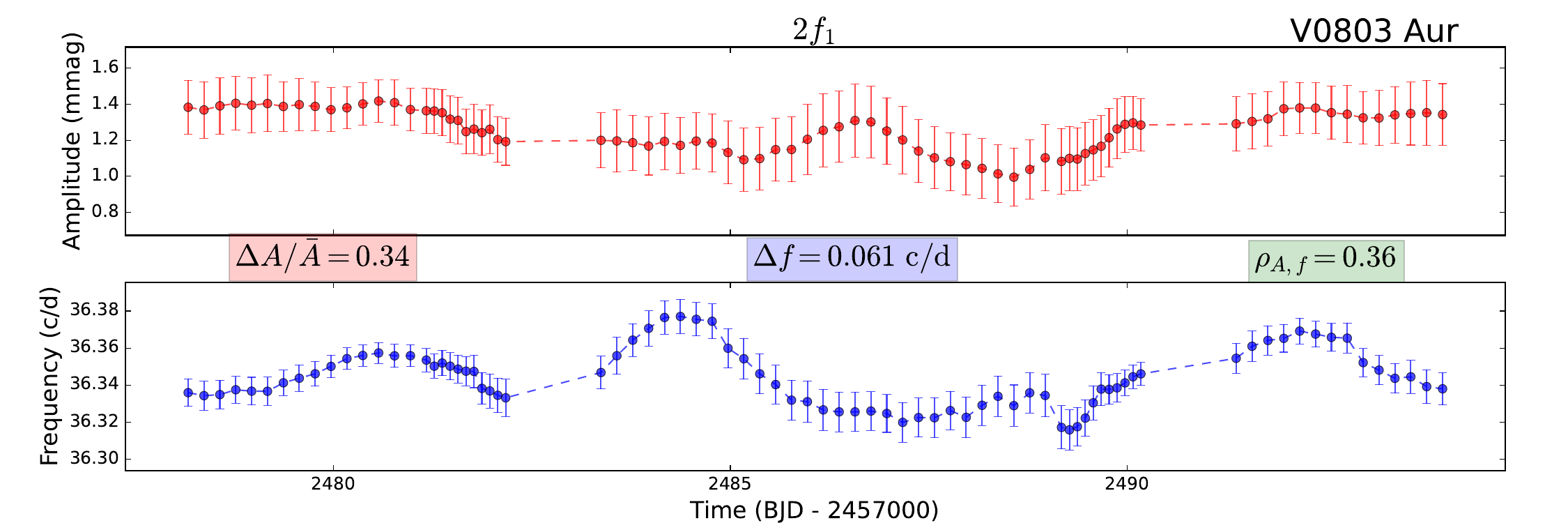}
  \includegraphics[width=0.48\textwidth]{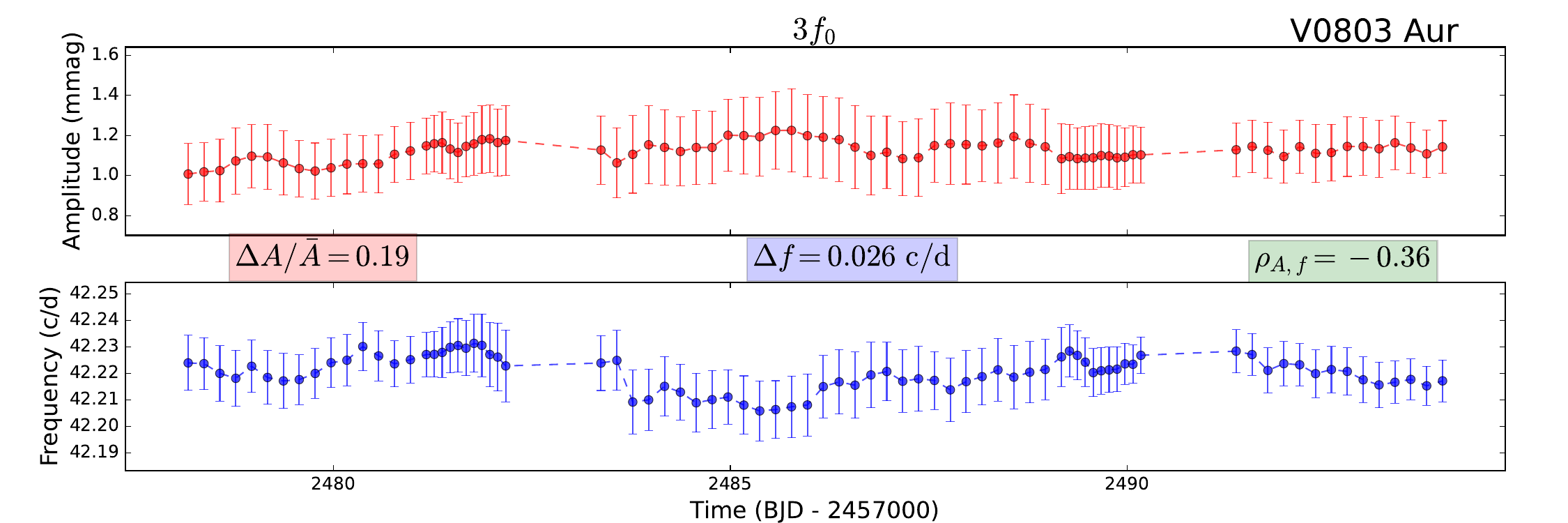}
  \phantom{\includegraphics[width=0.48\textwidth]{fig/V0803Aur_2f1.pdf}}
  \includegraphics[width=0.48\textwidth]{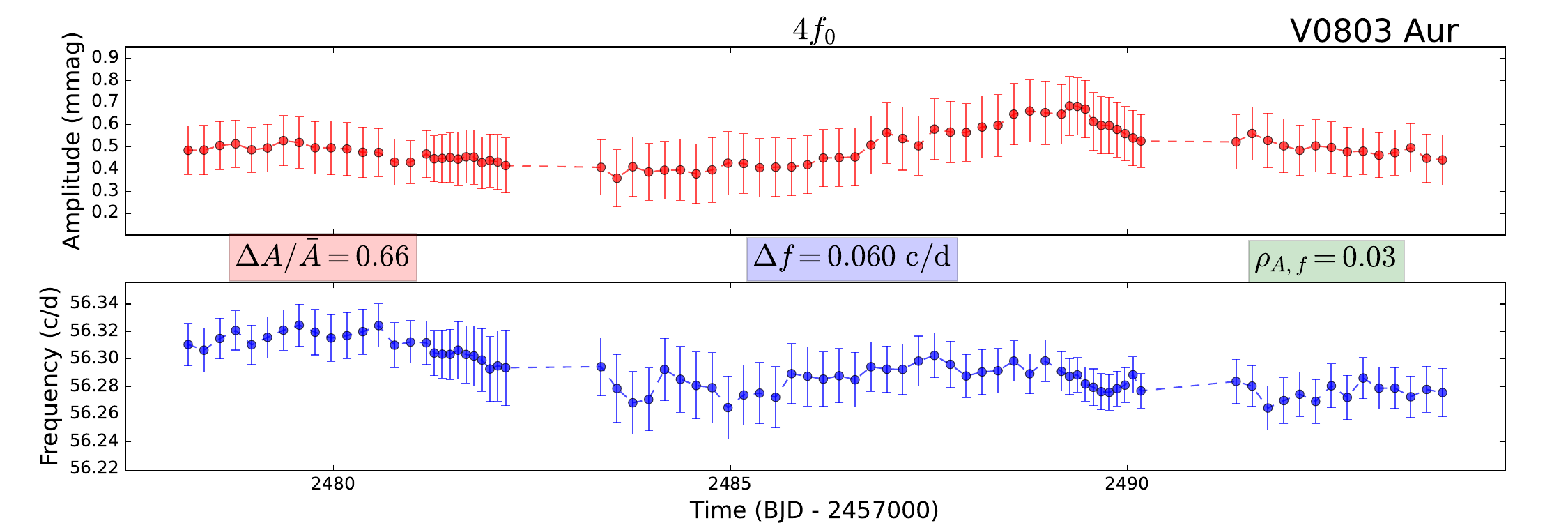}
  \phantom{\includegraphics[width=0.48\textwidth]{fig/V0803Aur_2f1.pdf}}
  \caption{Amplitude and frequency variations in the fundamental and first overtones and their harmonic pulsation modes ({$f_{0}$ through $4f_{0}$} and {$f_{1}$ through $2f_{1}$}) in V0803 Aur.}
  \label{fig:V0803Aur}
\end{figure*}

\begin{figure*}[htp]
  \centering
  \includegraphics[width=0.48\textwidth]{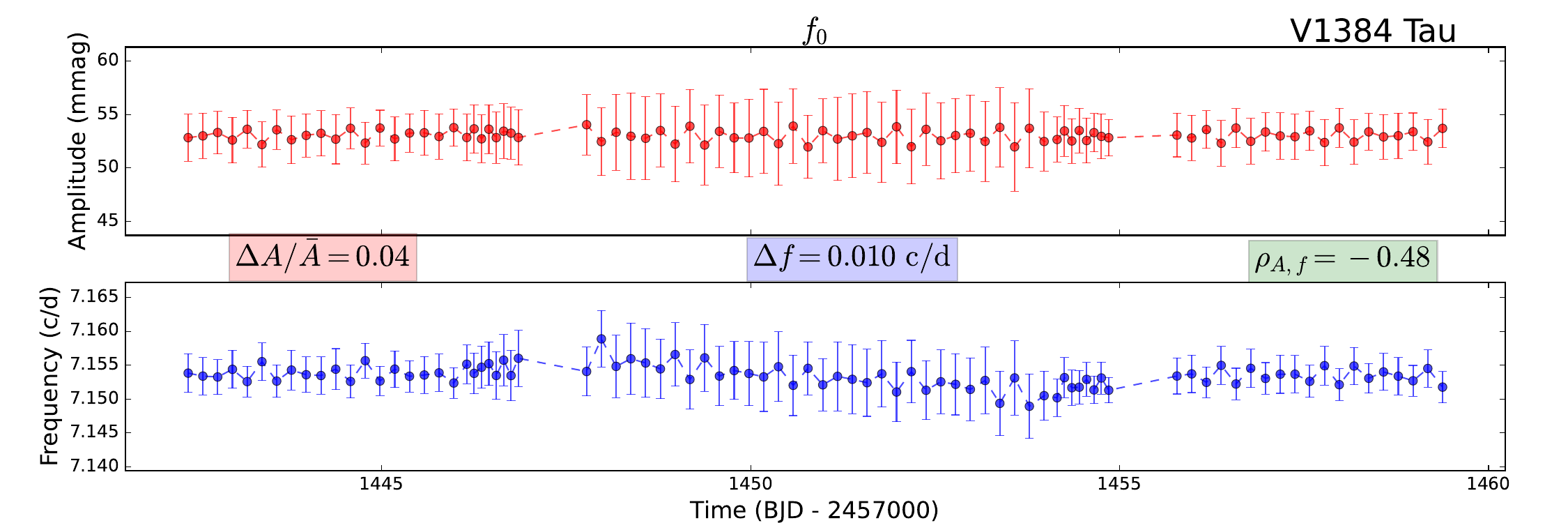}
  \includegraphics[width=0.48\textwidth]{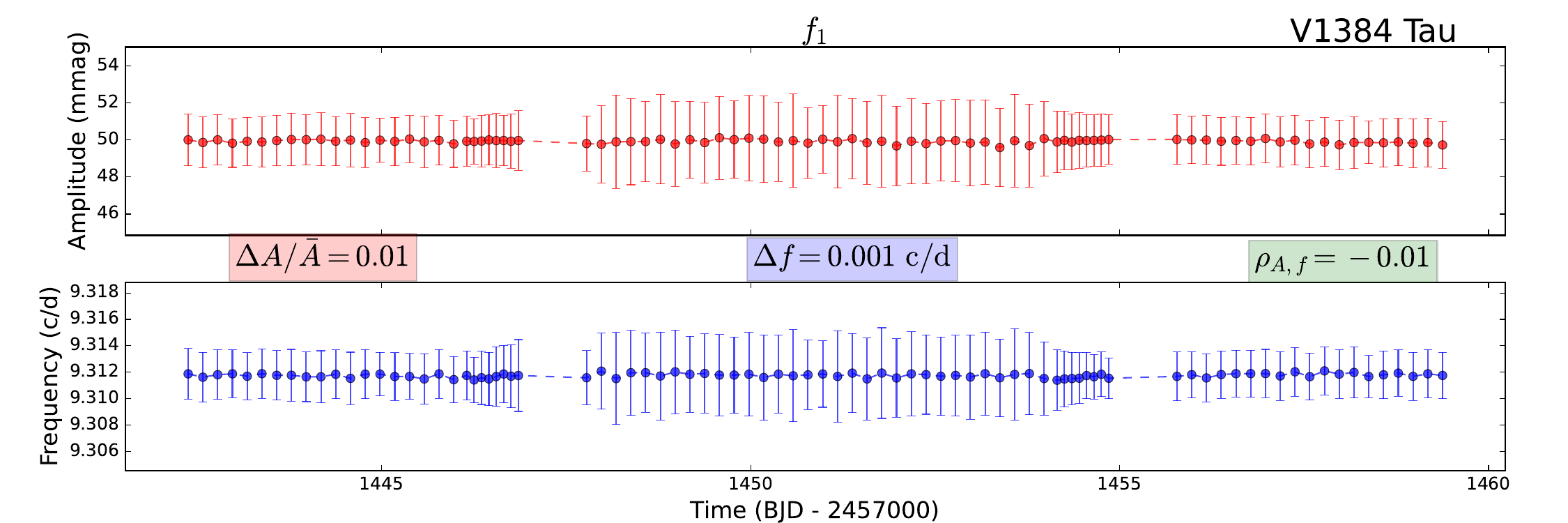}
  \includegraphics[width=0.48\textwidth]{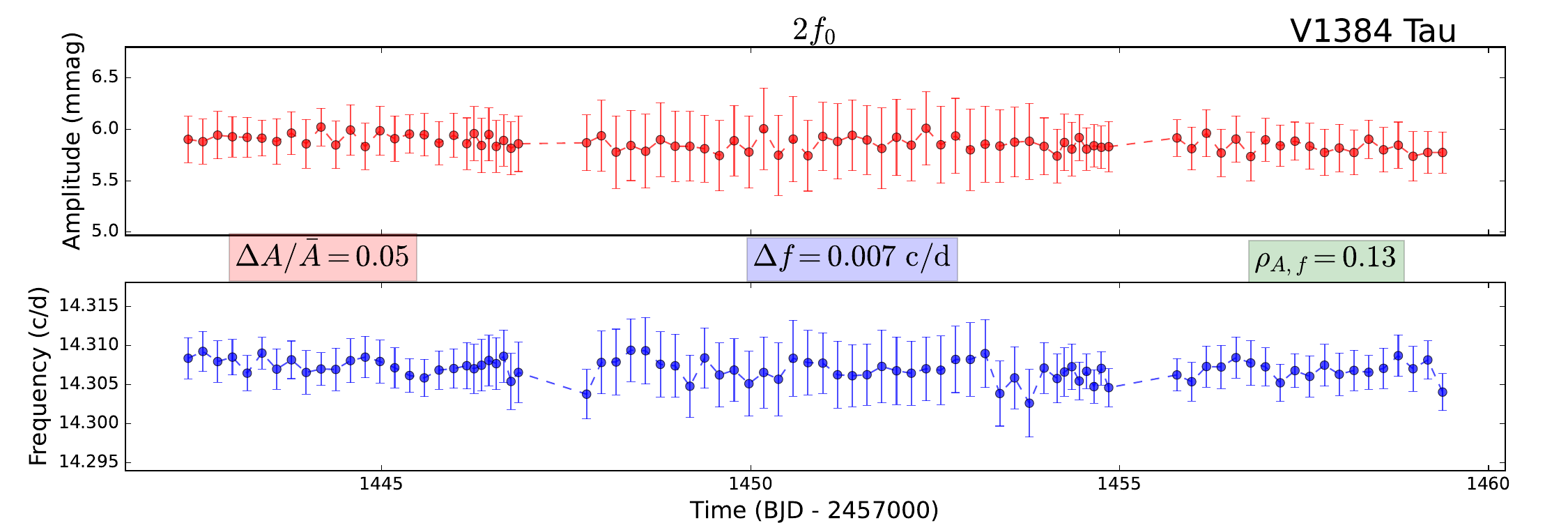}
  \includegraphics[width=0.48\textwidth]{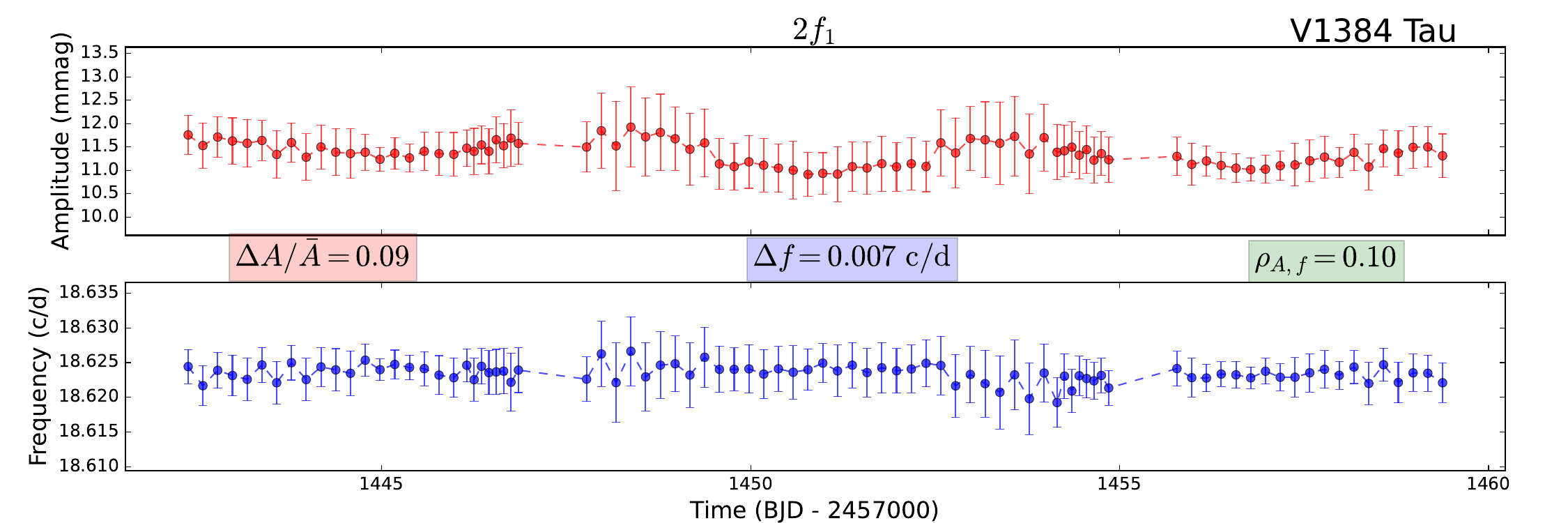}
  \includegraphics[width=0.48\textwidth]{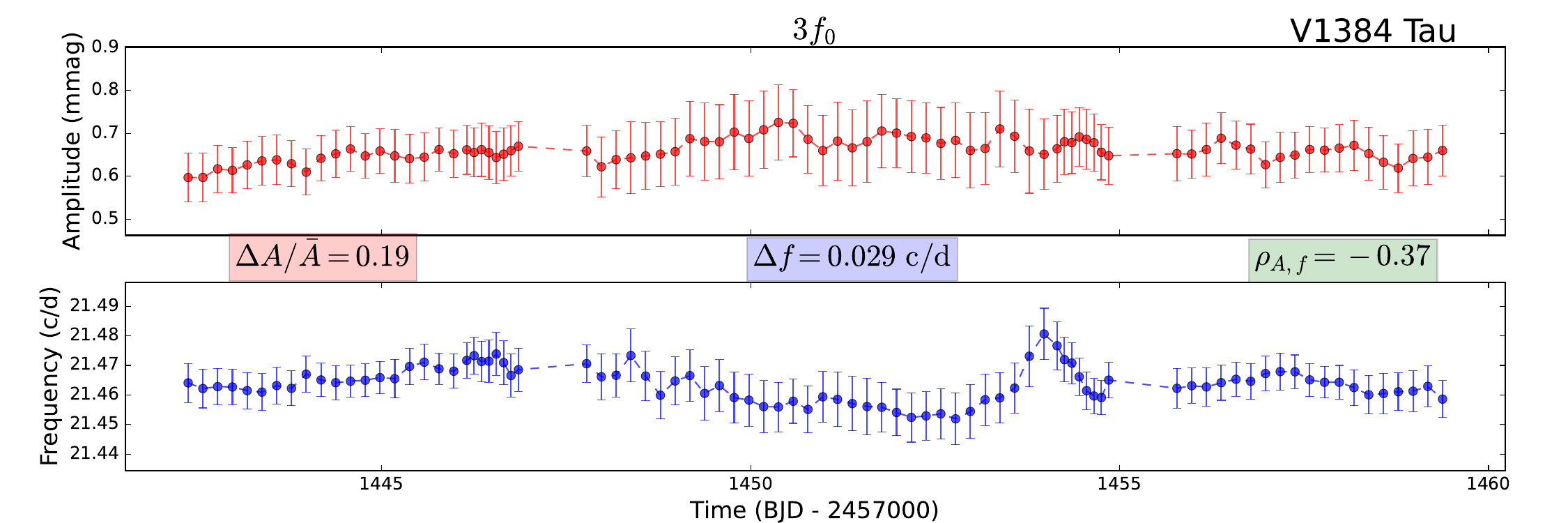}
  \includegraphics[width=0.48\textwidth]{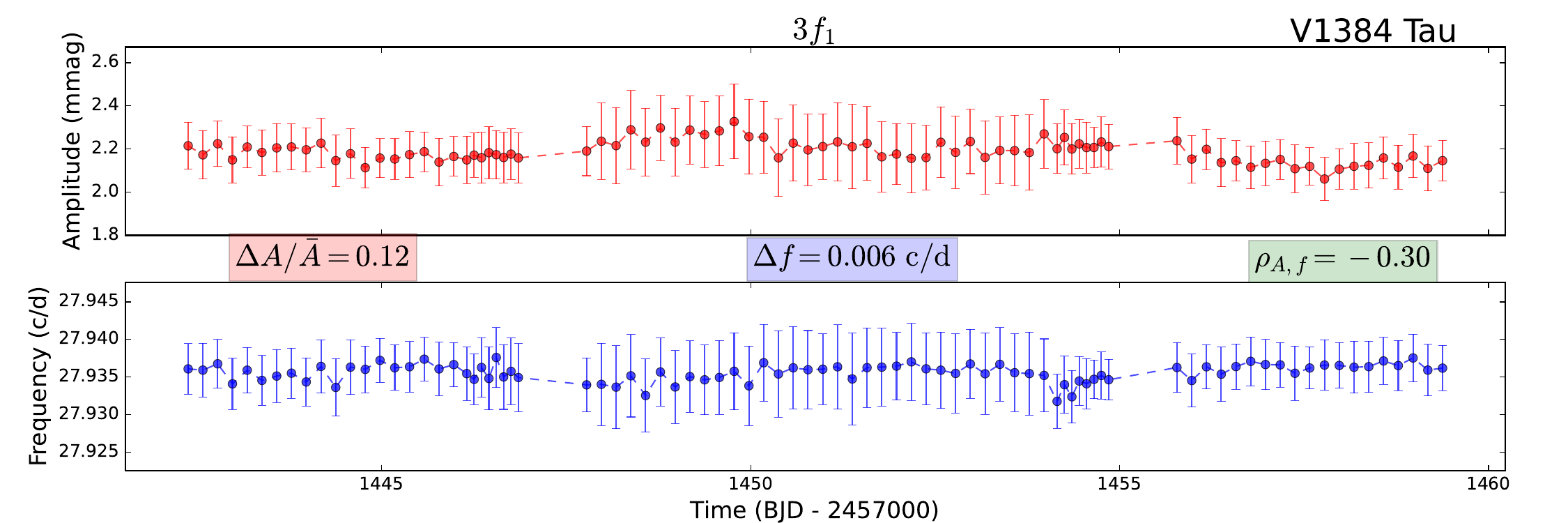}
  \caption{Amplitude and frequency variations in the fundamental and first overtones and their harmonic pulsation modes ({$f_{0}$ through $3f_{0}$} and {$f_{1}$ through $3f_{1}$}) in V1384 Tau.}
  \label{fig:V1384Tau}
\end{figure*}

\clearpage
\section{Interaction diagrams}
\label{app:IDs}

\begin{figure*}[htp]
  \centering
  \includegraphics[width=0.55\textwidth]{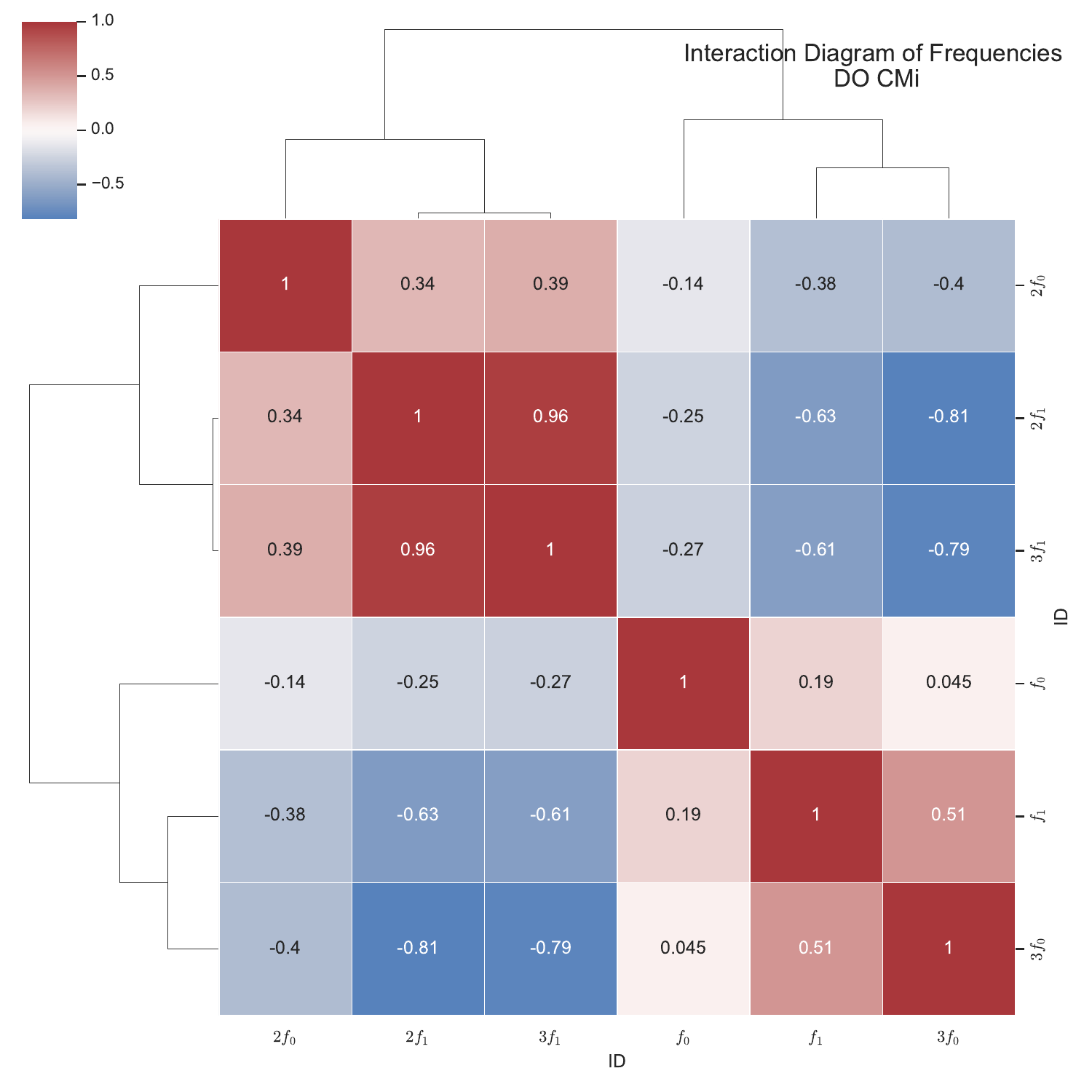}
  \caption{Interaction diagram of the frequencies of $f_0$ and $f_1$ and their harmonic pulsation modes in DO CMi.}
  \label{fig:DOCMi_ID}
\end{figure*}

\begin{figure*}[htp]
  \centering
  \includegraphics[width=0.55\textwidth]{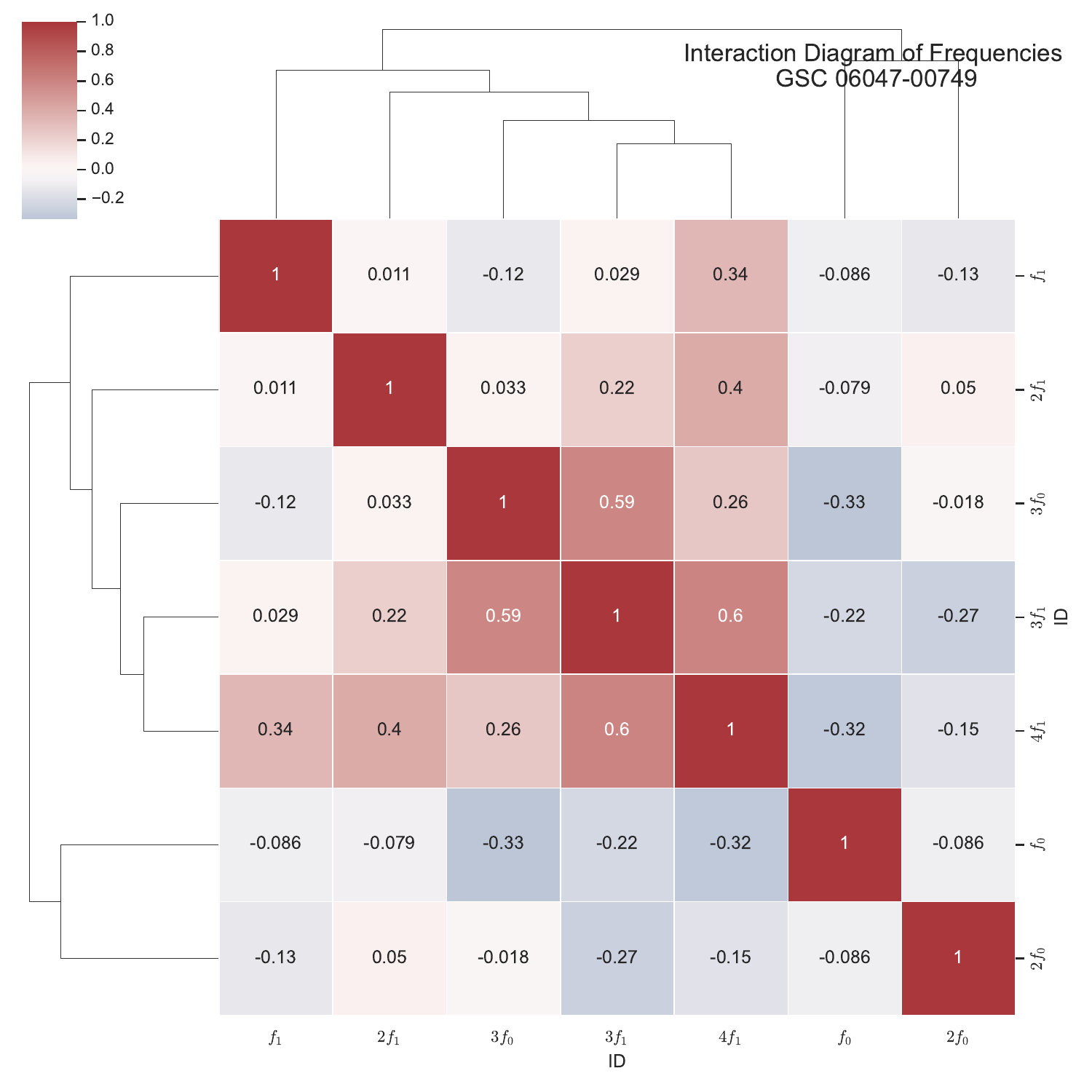}
  \caption{Interaction diagram of the frequencies of $f_0$ and $f_1$ and their harmonic pulsation modes in GSC 06047-00749.}
  \label{fig:GSC06047-00749_ID}
\end{figure*}

\begin{figure*}[htp]
  \centering
  \includegraphics[width=0.5\textwidth]{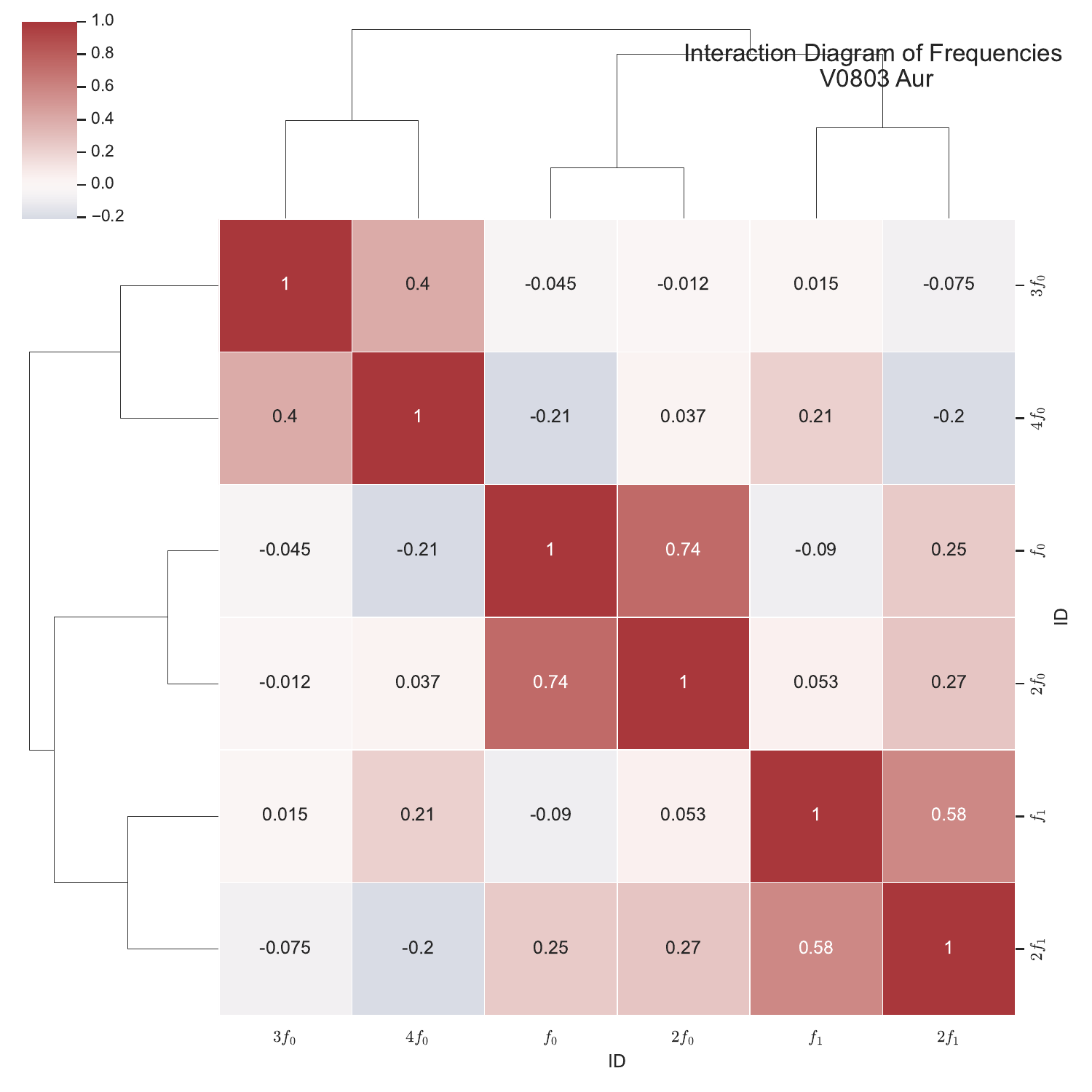}
  \caption{Interaction diagram of the frequencies of $f_0$ and $f_1$ and their harmonic pulsation modes in V0803 Aur.}
  \label{fig:V0803Aur_ID}
\end{figure*}

\begin{figure*}[htp]
  \centering
  \includegraphics[width=0.5\textwidth]{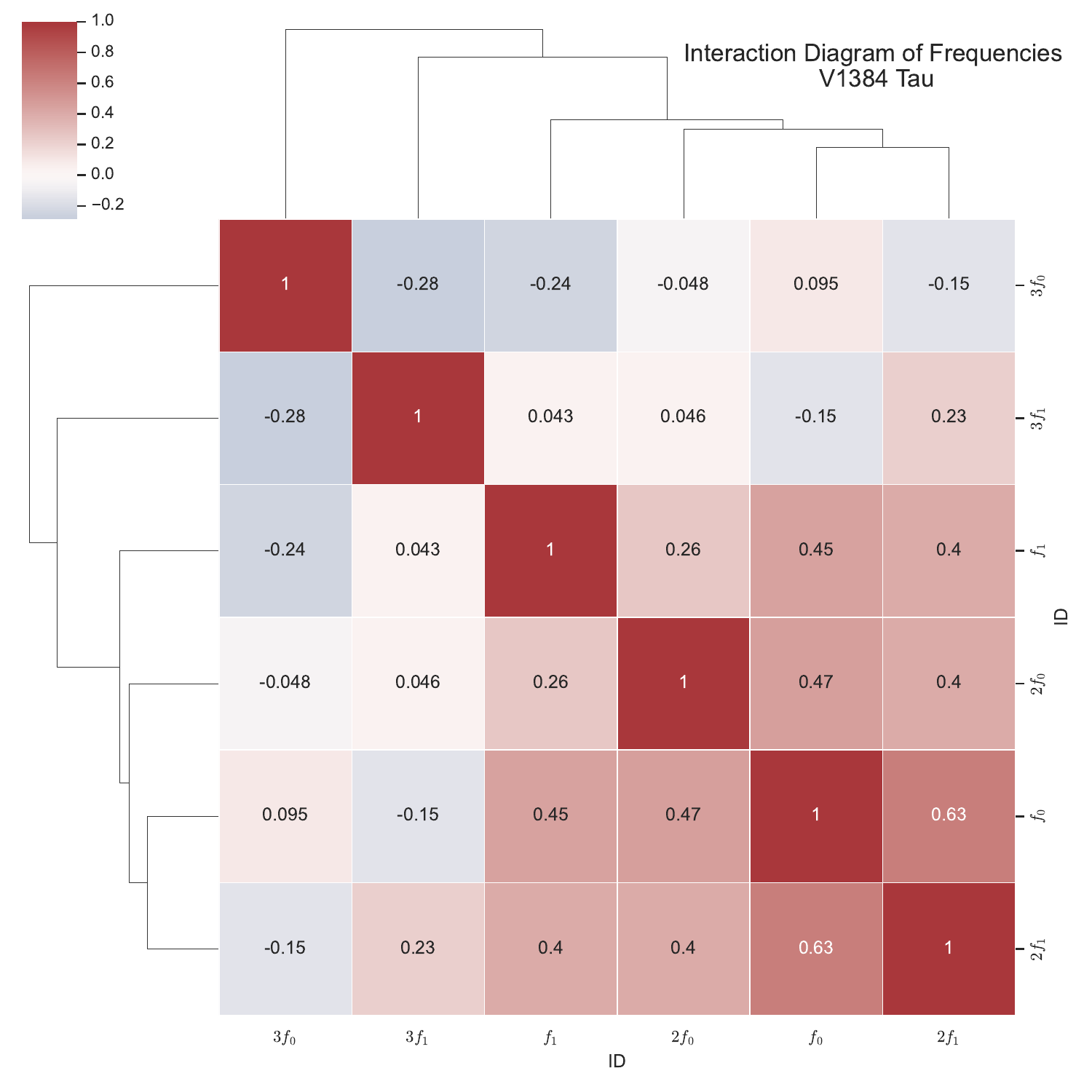}
  \caption{Interaction diagram of the frequencies of $f_0$ and $f_1$ and their harmonic pulsation modes in V1384 Tau.}
  \label{fig:V1384Tau_ID}
\end{figure*}

\section{Supplementary materials}
\label{app:SM}

\end{appendix}
\end{document}